\documentclass[%
 reprint,
 amsmath,amssymb,amsfonts
 aps,
]{revtex4-1}

\usepackage{natbib}
\usepackage{xcolor}

\usepackage{amsthm}   % Typesetting theorems (AMS style).

\theoremstyle{plain}% default

  \theoremstyle{definition}

\usepackage[titletoc,toc,title]{appendix}
\usepackage{graphicx}% Include figure files
\usepackage{dcolumn}% Align Table~columns on decimal point
\usepackage{bm}% bold math
\usepackage[utf8]{inputenc}
\usepackage{amsmath}
\usepackage{amsfonts}
\usepackage{mathtools}
\usepackage{dsfont}
\usepackage{float} % Allows putting an [H] in \begin{figure} to specify the exact location of the figure
\usepackage{wrapfig} % Allows in-line images such as the example fish picture
\usepackage[caption=false]{subfig}

\usepackage{hyperref}

\begin{document}

%\preprint{APS/123-QED}

\title{Conformally coupled tachyonic dark energy}% Force line breaks with \\

\author{Elsa M. Teixeira$^1$, Ana Nunes$^2$ and Nelson J. Nunes$^1$}
\affiliation{%
$^1$Instituto de Astrof\'{i}sica e Ci\^{e}ncias do Espa\c{c}o, Faculdade de 
Ci\^{e}ncias da Universidade de Lisboa, Campo Grande, PT1749-016 Lisboa, Portugal\\
$^2$BioISI - Biosystems and Integrative Sciences Institute, Faculdade de 
Ci\^{e}ncias da\\ Universidade de Lisboa, Campo Grande, PT1749-016 Lisboa, Portugal 
}%

\date{\today}% It is always \today, today,

\begin{abstract}
We present and study a conformally coupled dark energy model, characterised by an interaction between a tachyon field $\phi$, with an inverse square potential $V (\phi)$, and the matter sector. A detailed analysis of the cosmological outcome reveals different possibilities, in contrast with the previously studied uncoupled model, for which there exists only one stable critical point that gives late-time acceleration of the Universe. The introduction of the coupling translates into an energy exchange between the fluids. We find the interesting possibility of the emergence of a new fixed point, which is a scaling solution and an attractor of the system. In this case, it is possible to describe an everlasting Universe with $\Omega_{\phi} \simeq 0.7$, therefore alleviating the cosmic coincidence problem. However, we find that, in order for the evolution to be cosmologically viable, there is the need to introduce one uncoupled matter species, dominant at early times.
%\begin{description}
%\end{description}
\end{abstract}

\pacs{Valid PACS appear here}% PACS, the Physics and Astronomy
\maketitle

%\tableofcontents

\section{\label{sec:int}Introduction}

The discovery of the accelerated expansion of the Universe by the Supernova Cosmology Project \cite{acel2} and the High-z Supernova Search Team \cite{acel1} in 1998 resulted in drastic changes regarding our knowledge of the Universe. Based on several cosmological observations \cite{planck, planck2}, it is very well established that the Universe is currently undergoing a period of accelerated expansion. These results indicate that there seems to exist an unknown source of energy/matter component, which happens to be the most abundant among the known constituents of the Universe. Such an energy source would need to have an equation of state (EoS) parameter characterised by an effective negative pressure in order to explain the observations and is generally classified under the broad heading of dark energy (DE) \cite{amendola}.

The simplest model of dark energy is represented by the cosmological constant $\Lambda$, related to a cosmological source with a constant EoS parameter $ w_{\Lambda}= -1$, and is called the $\Lambda$CDM model \cite{carroll,Padmanabhan:2002ji}. CDM stands for a cold dark matter \cite{dm, dm2} component, needed to explain the measurements of the rotation curves of galaxies, which are not in agreement with the theoretical predictions based on Newtonian mechanics \cite{Bertone:2004pz}. 
However, according to the standard $\Lambda$CDM model, the Universe in the present day appears to be extremely fine-tuned, as this cosmological constant seems to have begun dominating the universal energy at a very specific moment. This is known as the cosmic coincidence problem and stands for one of the various conceptual problems related to the $\Lambda$CDM model \cite{padilla, weinberg2, cc2}. In attempts to avoid these problems, the cosmological constant is often generalised to a dynamical scalar field, whose time evolution could more naturally result in the energy density observed today \cite{cc1, Copeland:2006wr}.

Scalar field based theories of dark energy are most commonly described by a canonical scalar field $\varphi$, the \textit{quintessence} field \cite{tsuq, PhysRevD.58.023503, Barreiro:1999zs,  Chiba:1999ka, Copeland:1997et, delaMacorra:1999ff, Ng:2001hs, Bahamonde:2017ize, Albuquerque:2018ymr}. This is the simplest scalar field scenario, corresponding to a consistent field theory generalisation of the classical nonrelativistic particle framework, characterised by a Lagrangian density of the form:

\begin{equation}
\mathcal{L_{\rm quin}}=-\frac{1}{2} \partial_\mu \varphi\, \partial^\mu \varphi - V(\varphi).
\label{quin}
\end{equation}

Analogously, one could look for a natural scalar field generalisation of the Lagrangian for a relativistic particle. As addressed in Ref.~\cite{Padmanabhan:2002s}, this can be written as

\begin{equation}
\mathcal{L}_{\rm tach}=-V(\phi)\, \sqrt{1+\partial_{\mu} \phi\, \partial^{\mu} \phi},
\label{tach}
\end{equation}

\noindent where the field $\phi$ is termed the \textit{tachyon} field. In this manuscript we adopt the system of natural units, such that $c=\hbar=1$, for the speed of light and Planck's constant. In this framework the quintessence and tachyon scalar fields have dimensions of mass and mass$^{-1}$, respectively.

This relativistic approach includes the description of massless particles with a bounded well-defined (relativistic) energy. Also, it is possible to treat the mass as a function of the scalar field \cite{pad2}, through $V(\phi)$.

The tachyon scalar field is widely studied in the context of string theory, due to its crucial role in the Dirac-Born-Infeld (DBI) action, used to describe the D-brane action \cite{senroltac, sentacmat,senfieldteo,sencosmo, mohammadtac}. 
Tachyons were originally proposed as hypothetical particles which could travel faster than light. However, in the context of field theory, this has acquired new meaning and tachyons are associated with the description of quantum states with negative mass squared. This translates into a vacuum instability which is represented by the negative sign in Eq.~\eqref{tach}. The potential is initially at a local maximum, that is, the field is very precisely balanced at the top of the potential, and any small perturbation will destabilise the system and cause the field to roll towards the local minimum. By virtue of this process, the quanta become well-defined particles with a positive, real-valued mass. It has been shown that, in certain dynamical regimes (while not in general), an equivalence can be made between the tachyonic, Eq.~\eqref{tach}, and quintessence Lagrangians, Eq.~\eqref{quin} \cite{Malquarti:2003nn}.
One interesting feature is the fact that, regardless of the form of the potential, the equation of state parameter of the tachyon field is constrained to a range between $-1$ and $0$, and, therefore, phantomlike behaviour is not allowed for a tachyon field, with the Lagrangian density given in Eq.~\eqref{tach}.

It has already been shown that tachyons could play a useful role in cosmology \cite{tacgib, tacinf, Mukohyama, Sami, Mazumdar:2001mm, PhysRevD.67.063511, Das:2003xw, Gibbons:2003gb, NOJIRI2004137, PhysRevD.69.123517, Causse:2003hp, Guo:2004dta, Sami:2002zy, Gorini:2003wa, Frolov:2002rr}. 
Tachyons as a source of dark energy have also been the focus of many studies \cite{padd, ABRAMO2003165, Abramo:2003cp}. 
In Ref.~\cite{cop} a complete dynamical study of a dark energy scenario in the presence of a tachyon field and a barotropic perfect fluid for specific forms of the potential was performed. This was also done in Ref.~\cite{Fang:2010zze} for a general form of the potential. 
For the tachyon field, the simplest case corresponds to the inverse square potential, which was studied with dynamical system techniques, for example, in Refs.~\cite{cop, Aguirregabiria:2004xd}. 
Other forms for the potential were considered in Refs.~\cite{Quiros:2009mz, Guo:2003zf}.

The simplest dynamical DE models assume that the scalar field does not couple to the other matter forms. Taking purely phenomenological arguments, there is no reason to believe that this should be the case. Consequently, the simplest extension is to assume that the field is allowed to couple nonminimally to the matter sector (usually to the dark matter fluid) \cite{Ellis:1989as, Wetterich:1994bg, PhysRevD.48.3436}. From a dynamical system point of view, the advantage of introducing interacting dark energy models is the prospect of obtaining new scaling solutions in which the DE field shares with the matter sector a constant, nonvanishing fraction of the total energy density. The emergence of these solutions is a valuable feature of any model, as they can be employed in order to alleviate the coincidence problem. The possibility of an interactive dark energy scalar field coupled to a matter component and its respective cosmological consequences were seminally discussed in Refs.~\cite{Ellis:1989as, Wetterich:1994bg, PhysRevD.48.3436, DAMOUR1994532}. 
The so-called coupled quintessence models were first introduced and analysed in Refs.~\cite{PhysRevD.62.043511,  Zimdahl:2001ar} as an extension of nonminimally coupled theories \cite{Amendola:1999qq}. A variety of different models were soon proposed and are distinguishable by the form of the coupling \cite{Billyard:2000bh, Holden:1999hm, Amendola:2000uh, TocchiniValentini:2001ty, Gonzalez:2006cj, PhysRevD.73.023502, Boehmer:2008av, Boehmer:2009tk,Wei:2010fz, Xu:2012jf, Hossain:2014xha, Shahalam:2015sja, nelson, Barros:2018efl, Frusciante:2018aew,  Barros:2019rdv}.
Coupled tachyonic models have also been considered for different coupling functions \cite{Gumjudpai:2005ry, Farajollahi:2011jr, Farajollahi:2011ym, Landim:2015poa, Shahalam:2017fqt, Ahmad:2015sna, Herrera:2004dh}.

Most commonly, the coupling function is taken to depend upon parameters of the model such as the Hubble rate $H$, the scale factor $a$, the fluid and/or field's energy density $\rho$, and the field $\phi$ and/or its derivatives, and it is imposed at the level of the field equations, with its cosmological implications being studied afterward. However, the coupling could be more naturally generated. One way to introduce a nontrivial coupling between the scalar field and matter is to envision that matter particles follow geodesics defined with respect to a transformed metric $\widetilde{g}_{\mu \nu} $ which is related to the gravitational metric $g_{\mu \nu}$ by means of a conformal transformation, casting the interaction into a Lagrangian description \cite{Faraoni:1998qx}.

In this work, we propose a tachyonic dark energy model in which the DE field is allowed to couple to the matter sector by means of a conformal transformation of the metric. The evolution of the DE scalar field under this model is characterised by an inverse square potential and a power-law coupling function. We study how the introduction of the coupling affects the dynamics of the model and the corresponding cosmological outcome, in comparison with the previously studied uncoupled tachyonic dark energy model \cite{cop, padd, Fang:2010zze}. In the latter, there exists only one stable critical point capable of describing the late-time acceleration of the Universe, corresponding to a totally dark energy dominated future configuration. The conformally coupled model, by contrast, provides a scaling solution, allowing for different regimes \cite{Sen:2008yt}. Based on the dynamical analysis, we conclude that this model is capable of reproducing the cosmic history of the Universe.

The paper is structured as follows: In Secs. \ref{sec:mod} and \ref{sec:bac} the cosmological model is introduced and the background equations are derived for a spatially flat Friedmann-Lema\^{i}tre-Robertson-Walker (FLRW) Universe. Then, in Sec. \ref{sec:dyn}, we derive the equations of motion in terms of some convenient dynamical variables. Additionally, we provide a detailed study of the evolution of the system, as a function of the parameters. This allows for the investigation of viable cosmologies and for a detailed analysis of the cosmological predictions of the model, in direct comparison with the purely uncoupled case. This study is summarised in Sec. \ref{sec:vc}. Additionally, in Sec. \ref{sec:eff}, we comment on the interpretation of the presence of the coupling as a contribution to an effective potential for the theory. Finally, we conclude in Sec. \ref{sec:conc}.

\section{\label{sec:mod}The Model}

Let us consider a four-dimensional spacetime manifold $\mathcal{M}$ endowed with a metric $g_{\mu \nu}$. We are interested in studying how a conformal coupling between dark energy and matter can affect the dynamics of the Universe. To do so, from now on we assume the possibility of having conformal transformations of the metric from the Einstein frame to the Jordan frame,
\begin{equation}
g_{\mu \nu}^i \longmapsto \widetilde{g}^{\,i}_{\mu \nu} =C_i(\phi)\, g_{\mu \nu} ,
\label{trans}
\end{equation}
\noindent where $C_i(\phi)$ are the conformal coupling functions and a tilde denotes quantities in the Jordan frame. Hereafter we will simply write $C_i \equiv C_i(\phi)$.

We find
\begin{equation}
g^{\mu \nu} \longmapsto \widetilde{g_i}^{\mu \nu} = \frac{1}{C_i}\, g^{\mu \nu}
\label{inverse}
\end{equation}
\noindent for the inverse of the conformal metric, Eq.~\eqref{trans}, and
\begin{equation}
g \equiv \det(g_{\alpha \beta}) \longmapsto \widetilde{g}_i \equiv \det(\widetilde{g}^{\,i}_{\alpha \beta}) = C_i^{\,4}\,  g,
\label{detmc}
\end{equation}
\noindent for the determinant of the metric.

We suppose that the field $\phi$ in the conformal transformation, Eq.~\eqref{trans}, is the one portraying dark energy, and therefore the coupling between matter and dark energy is fully accounted for if we consider that the gravitational Lagrangian depends on the metric $g_{\mu \nu}$, and that the matter Lagrangian is a function of the conformal metric $\widetilde{g}^i_{\mu \nu}$ defined in Eq.~\eqref{trans}. For this purpose, we consider a scalar-tensor theory in the Einstein frame with action

\begin{equation}
\begin{aligned}
S=&\int d^4x \sqrt{-g} \left[\frac{M_{P}^2}{2} R + \mathcal{L}_{\phi}(\phi, X)\right] \\
 + & \sum_i \int d^4x \sqrt{-\widetilde{g_i}} \ \widetilde{\mathcal{L}}_i (\widetilde{g}^{\,i}_{\mu \nu}, \psi_i, \partial_{\mu} \psi_i),
\end{aligned}
\label{action}
\end{equation}

\noindent where $M_P\equiv 1/ \sqrt{8 \pi G}$ is the reduced Planck mass and $R$ is the Ricci scalar, which depicts the geometrical sector of the Universe and, therefore, is defined in terms of $g_{\mu \nu}$. The term $\mathcal{L}_{\phi} $ stands for the Lagrangian density of the scalar field (associated with dark energy). Finally, $\widetilde{\mathcal{L}}_i$ is the Lagrangian for each of the matter fluids, allowed to depend on $\psi_i$ and its derivatives, where $\psi_i$ denotes matter fields propagating on geodesics defined by the metrics $\widetilde{g}^i_{\mu \nu}$. 

Throughout this chapter, we shall consider that the role of dark energy is played by a tachyon scalar field. This means that we take the tachyonic Lagrangian given in Eq.~\eqref{tach}, which may equivalently be written as
\begin{equation}
\mathcal{L}_{\phi}  = -V(\phi)\, \sqrt{1-2X},
\label{ptach}
\end{equation}
\noindent where $X=-\frac{1}{2} g^{\mu \nu} \partial_{\mu} \phi \partial_{\nu} \phi$ is the kinetic term associated with the tachyon scalar field $\phi$, and $V(\phi)$ is a general self-coupling potential. It can immediately be noted that the assumption $1-2X \geq 0$ is needed in order to assure one that the Lagrangian is real valued.

Variation of the action in Eq.~\eqref{action} with respect to the metric $g_{\mu \nu}$ leads to the Einstein field equations in the Einstein frame
\begin{equation}
G_{\mu \nu} \equiv R_{\mu \nu} - \frac{1}{2} g_{\mu \nu} R = \kappa^2 \left(T_{\mu \nu}^{\, \phi}  + \sum_i T^{\, i}_{\mu \nu}  \right),
\label{efeq}
\end{equation}
\noindent where $\kappa \equiv M_P^{-1} $ is the scaled gravitational constant, $G_{\mu \nu}$ is the Einstein tensor, $R_{\mu \nu}$ is the Ricci curvature tensor, both computed from $g_{\mu \nu}$, and $T_{\mu \nu}^{\phi}$ and $T^i_{\mu \nu}$ are the energy-momentum tensors for the scalar field $\phi$ and the matter fluids, defined, respectively, as
\begin{equation}
T_{\mu \nu}^{\, \phi} \equiv -\frac{2}{\sqrt{-g}} \frac{\delta \Big(\sqrt{-g}\, \mathcal{L}_{\phi}\Big)}{\delta g^{\mu \nu}}, \  \ \ T^{\,i}_{\mu \nu} \equiv- \frac{2}{\sqrt{-g}} \frac{\delta \Big(\sqrt{-\widetilde{g}_i}\, \widetilde{\mathcal{L}}_i \Big)}{\delta g_i^{\mu \nu}} .
\label{Tc}
\end{equation} 
\noindent From the expression of the tachyonic Lagrangian density, Eq.~\eqref{ptach}, and the previous definition, we gather that the energy-momentum tensor for the tachyon field reads
\begin{equation}
T_{\mu \nu}^{\, \phi}= V(\phi)\, \frac{ \partial_{\mu} \phi\, \partial_{\nu} \phi}{\sqrt{1+\partial^{\alpha} \phi\, \partial_{\alpha} \phi}} - V(\phi)\, g_{\mu \nu}\,  \sqrt{1+\partial^{\alpha} \phi\, \partial_{\alpha} \phi}.
\label{ttac}
\end{equation}  

The Einstein tensor $G_{\mu \nu}$ is divergenceless, $\nabla^{\mu} G_{\mu \nu}=0$, and from this relation it follows that
\begin{equation}
\nabla^{\mu} \left(T_{\mu \nu}^{\, \phi} + \sum_i T^{\, i}_{\mu \nu}  \right)=0.
\label{emocons}
\end{equation} 
\noindent As expected, since the $\widetilde{\mathcal{L}}_i$ terms depend on the field $\phi$ (through $\widetilde{g}^i_{\mu \nu}$), the energy-momentum tensors of dark energy and coupled matter fluids are not separately conserved. 

The energy-momentum tensor in the Einstein frame, as defined in Eq.~\eqref{Tc}, is related to the one in the Jordan frame as
\begin{equation}
T_i^{\, \alpha \beta}= \frac{\sqrt{-\widetilde{g}_i}}{\sqrt{-g}}\,  \frac{\delta\widetilde{g}^{\, i}_{\rho \sigma}}{\delta g_{\alpha \beta}}\, \widetilde{T}_i^{\, \rho \sigma},
\end{equation}
\noindent where $ \sqrt{-\widetilde{g}_i}/\sqrt{-g}$ is the Jacobian of the transformation, defined in terms of the determinants of each metric, related by Eq.~\eqref{detmc}.

The \textit{coupled equation of motion} for the scalar field is computed from variation of the action in Eq.~ \eqref{action} with respect to $\phi$ (again recalling that $\widetilde{\mathcal{L}}_i$ depends on $\phi$ through $\widetilde{g}_{\mu \nu}^{i}$):
\begin{equation}
\Box \phi - \frac{\nabla^{\mu}\, \partial^{\nu} \phi\, \partial_{\mu} \phi\, \partial_{\nu} \phi}{1+\partial^{\mu} \phi\, \partial_{\mu} \phi} -\frac{V_{, \phi}}{V} =-  \frac{\sqrt{1+\partial^{\mu} \phi\, \partial_{\mu} \phi}}{V}\, \sum_i Q_i,
\label{eqmotach}
\end{equation}
\noindent where $\Box=g^{\mu \nu} \nabla_{\mu} \nabla_{\nu}$ is the d'Alembert operator, and
\begin{equation}
Q_i=\frac{1}{2}  \frac{\sqrt{-\widetilde{g}_i}}{\sqrt{-g}}  \left({C_i}_{, \phi}\, g_{\mu \nu} \widetilde{T}_i^{\, \mu \nu} \right)=\frac{{C_i}_{, \phi} }{2 C_i}\, T_i
\label{Q}
\end{equation}
\noindent is the interaction term associated with the $i$th coupled fluid, where $T_i \equiv g_{\alpha \beta} T_i^{\alpha \beta}$ is the trace of the energy-momentum tensor $T_i^{\mu \nu}$. Even though the energy-momentum tensor for each fluid is not independently conserved, according to Eqs.~\eqref{ttac}, \eqref{emocons}, and \eqref{eqmotach}, we find the conservation relations
\begin{equation}
\nabla^{\mu} T_{\mu \nu}^{\, \phi}= - \sum_i Q_i\, \partial_{\nu} \phi
\label{qna}
\end{equation}
\noindent and
\begin{equation}
\nabla^{\mu} T^{\, i}_{\mu \nu}= Q_i\, \partial_{\nu} \phi.
\label{qnab}
\end{equation}

\section{\label{sec:bac}Background Cosmology}

For cosmological applications, we consider the homogeneous, isotropic, spatially flat FLRW Einstein frame metric defined according to
\begin{equation}
ds^2= g_{\mu \nu}\, dx^{\, \mu} dx^{\, \nu} =- dt^2+a^2(t)\, \delta_{ij}\, dx^{\, i} dx^{\, j},
\label{frwmetricc}
\end{equation}
\noindent where $t$ is the cosmic time, and $a(t) > 0$ is the scale factor which describes the expansion of the Universe [defined in a way such that $a(t_{\text{today}})=1$]. 

We assume energy.momentum tensors of a perfect fluid form for each of the $i$ components:
\begin{equation}
T^{\, i}_{\mu \nu} = (\rho_i + p_i)\, u_{\mu}\, u_{\nu} + p_i\, g_{\mu \nu},
\label{tperf}
\end{equation}
\noindent where $\rho_i$ and $p_i$ are, respectively, the Einstein frame energy density and pressure for the $i$th matter fluid, and $u_{\mu}$ is the fluid's four-velocity for a comoving observer, which, by definition, satisfies $u^{\mu} u_{\mu}=-1$.

The energy-momentum tensor of the field can also be thought of as having a perfect fluid form,
\begin{equation}
T_{\mu \nu}^{\, \phi} = (\rho_{\phi} + p_{\phi})\, u_{\mu}\, u_{\nu} + p_{\phi}\, g_{\mu \nu},
\label{tperff}
\end{equation}
 \noindent provided that, and according to Eq.~\eqref{ttac}, the field's energy density and pressure are written as
\begin{equation}
\begin{aligned}
\rho_{\phi}=\frac{V(\phi)}{\sqrt{1-\dot{\phi}^2}}\ \ \ \text{and}\ \ \  p_{\phi}=-V(\phi)\, \sqrt{1-\dot{\phi}^2},
\end{aligned}
\label{rphi}
\end{equation}
\noindent where, analogously, $\rho_{\phi}$ and $p_{\phi}$ are the energy density and pressure for the field. 
Directly from Eq.~\eqref{rphi} we gather that
\begin{equation}
w_{\phi} = \frac{p_\phi}{\rho_\phi}=\dot{\phi}^2-1,
\end{equation}
\noindent for the EoS parameter of the field. Note that the assumption $1-2X\geq0$ now translates into $\dot{\phi}^2\leq 1$. Consequently, the tachyon field presents a very particular type of dynamics in the sense that, regardless of the steepness of the potential, its equation of state parameter always varies between $0$, in which case it behaves like a dustlike fluid, and $-1$, resembling a cosmological constant. Also, this model cannot feature fields with a phantom nature \cite{amendola}, since that would imply that $w_\phi<-1$. In order to satisfy the current cosmological constraints pointing to $w_{\phi} \approx -1$, the kinetic energy of the tachyon scalar field must be practically negligible at late times, i.e., $\dot{\phi}^2 \ll 1$.
We directly conclude that the tachyon energy density evolves according to $\rho_{\phi} \propto a^{-p}$ with $0 \leq p \leq 3$. 

The dynamical equations arising from Eqs.~\eqref{eqmotach} and \eqref{qnab}, together with Eq.~\eqref{tperf}, yield the coupled equation of motion and the fluid conservation equations:
\begin{equation}
\ddot{\phi} + \left (1-\dot{\phi}^2 \right) \left(3 H \dot{\phi} + \frac{V_{, \phi}}{V}  \right)= \frac{1}{V} \left(1-\dot{\phi}^2 \right)^{3/2}\, \sum_i Q_i,
\label{kg}
\end{equation}
\begin{equation}
\dot{\rho}_i + 3 H \rho_i \left( 1+ w_i \right) = - Q_i\, \dot{\phi} ,
\label{fluidx}
\end{equation}
\noindent where an upper dot denotes time derivatives and $H (t) \equiv \dot{a}/a$ is the Hubble rate, whose dynamical evolution is determined by the Friedmann equations, computed directly from the Einstein field equations, together with Eqs.~\eqref{tperf} and \eqref{tperff}:
\begin{equation}
H^2= \frac{\kappa^2}{3} \left(\rho_{\phi}+ \sum_i \rho_i \right),
\label{fried1} 
\end{equation}
\begin{equation}
\dot{H}=- \frac{\kappa^2}{2} \left(\rho_{\phi} \left(1+w_{\phi} \right) + \sum_i\rho_i \left(1+ w_i \right)  \right).
\label{fried2}
\end{equation}
For cosmological purposes, we are interested in the Einstein frame dynamics of the scalar field in the presence of radiation and matter. Therefore, we consider a matter sector enclosing effective nonrelativistic and relativistic perfect fluids. Note that, in this model, the relativistic fluids are always conformally invariant and, therefore, do not couple to the dark energy fluid. As a first approximation, we ignore the contribution coming from the baryons, due to the stringent constraints from Solar System bounds \cite{Wang:2016lxa}. This is taken into account in most of the works present in the literature, on the basis of dark matter standing for the majority of the pressureless matter sourcing the right-hand side (rhs) of the Einstein equations.
Taking this into account, it is possible to rewrite the conservation relations in Eq.~\eqref{fluidx} as
\begin{equation}
\dot{\rho_{ r}} + 4 H \rho_{r}  = 0 ,
\label{rcon}
\end{equation}
\begin{equation}
\dot{\rho}_m + 3 H \rho_m = - Q\, \dot{\phi},
\label{fluidcon}
\end{equation}
\begin{equation}
\dot{\rho}_{\phi} + 3 H \rho_{\phi} (1+ w_{\phi}) =  Q\, \dot{\phi} ,
\label{fieldcon}
\end{equation}
\noindent where the subscripts $({m})$ and $({r})$ denote matter and radiation fluids, respectively, and Eq.~\eqref{fieldcon} is simply derived from Eqs.~\eqref{kg} and \eqref{qna}. 
$Q$ is the interaction term introduced in Eq.~\eqref{Q} for a pressureless matter fluid, i.e., for $w_m=0$ and $C_m (\phi) \equiv C$, some arbitrary function of the $\phi$ field, which reads
\begin{equation}
Q= -\frac{C_{,\phi}}{2 C}\, \rho_{m}.
\label{Qfrwq}
\end{equation}
 This is the term promoting the exchange of energy between the species.
From Eqs.~\eqref{fluidcon} and \eqref{fieldcon}, we infer that, whenever $Q \dot{\phi} >0$, energy is being transferred from the matter sector to dark energy, whereas when $Q \dot{\phi}<0$, it is the dark energy fluid which concedes energy to the matter fields. 

\begin{table}[t]
\centering
\caption[Couplings of a barotropic perfect fluid to a tachyonic field studied in the literature]{Couplings of a barotropic nonrelativistic perfect fluid to a tachyonic dark energy component studied in the literature, with $Q$ as defined in Eqs.~\eqref{kg} and \eqref{fluidx}. Note that $\rho$, $\rho_m$, and $\rho_{\phi}$ stand for the energy density of a general barotropic perfect fluid, a pressureless fluid, and the dark energy field, respectively, and $q$ stands for a constant, $C \equiv C(\phi)$, and $f \equiv f(\phi)$.}
\begin{tabular}{ l c }
\hline\hline \\[-2ex]
Reference & $Q$\\
\hline \\[-1.5ex]
Gumjudpai \textit{et. al} (2005) \cite{Gumjudpai:2005ry}  & $-q \rho $\\
Farajollahi and Salehi (2011) \cite{Farajollahi:2011jr}  & $\ \ \ - \frac{f_{, \phi}}{f} \rho (1-3w)\ $\\
Landim (2015) \cite{Landim:2015poa} & $-\frac{ q}{H} \rho_m \rho_{\phi}$\\
Shahalam \textit{et. al} (2017) \cite{Shahalam:2017fqt} & $- q  \dot{\rho_{\phi}} / \dot{{\phi}} $\\
This work & $-\frac{C_{,\phi}}{2C} \rho (1-3 w)$\\[1.5ex]
\hline
\end{tabular}
\label{table:coup}
\end{table}

Other forms for the coupling function $Q$, as defined in Eqs.~\eqref{kg} and \eqref{fluidx}, have been considered in the literature and are listed in Table~\ref{table:coup}, where $q$ stands for a constant, $\rho$, $\rho_m$, and $\rho_{\phi}$ stand for the energy density of a general barotropic perfect fluid, a pressureless fluid, and the dark energy field, respectively. $C \equiv C(\phi)$ and $f \equiv f(\phi)$ are free functions of the scalar field (defined at the level of the action). Inspection of Table~\ref{table:coup} leads to the conclusion that the general form of the coupling derived in this work is the same as the one considered in Ref.~\cite{Farajollahi:2011jr}. However, the model we study here is fundamentally different because of the different choices made for the form of the functions $f$ and $C$. This translates into a different set of fixed points and a new phenomenology.

\section{\label{sec:dyn}Dynamical System}

In order to study the evolution of the Universe under this model, we reduce the above system of Eqs.~\eqref{kg}--\eqref{fried2} and \eqref{Qfrwq} to a set of first order autonomous differential equations.
To do so, we introduce the following dimensionless variables, generalising the ones introduced in Ref.~\cite{cop}: 
\begin{equation}
\begin{aligned}
& x\equiv \dot{\phi},\ \ \  y^2\equiv\frac{\kappa^2 V}{3 H^2},\ \ \  z^2\equiv \frac{\kappa^2 \rho_{m}}{3 H^2},\ \ \ r^2\equiv \frac{\kappa^2 \rho_{r}}{3 H^2},\\
&\lambda\equiv - \frac{1}{\kappa}\, \frac{V_{, \phi}}{V^{3/2}},\ \ \ \sigma \equiv - \frac{1}{\kappa}\, \frac{C_{, \phi}}{C V^{1/2}},
\end{aligned}
\label{variables}
\end{equation}
\noindent where $\lambda$ stands for the steepness of the potential function and $\sigma$ relates the form of the conformal coupling function with the definition of the potential. For what concerns this work, we will consider $\lambda$ and $\sigma$ to be constants, i.e., the only free parameters of the model. This choice corresponds to a scalar field potential, $V$, and a conformal coupling function, $C$, of the form
\begin{equation}
\begin{aligned}
&V(\phi) =  \frac{V_0^2}{\phi^2}, \ \ \text{for}\  \ \lambda \neq 0 \ \  \text{\rm and} \\
 &C(\phi)=  \left(\frac{\phi}{\phi_0} \right)^{\frac{-2 \sigma}{\lambda}},
 \end{aligned}
\label{cons}
\end{equation}
\noindent where $V_0\equiv 2/(\kappa \lambda)$ is a mass scale associated with the scalar potential. Also, note that for tachyonic dark energy, the simplest closed system of autonomous equations is the one characterised by the inverse square potential \cite{Aguirregabiria:2004xd,cop} and not by the exponential potential, in contrast with canonical quintessence \cite{Copeland:1997et, Ng:2001hs, Barreiro:1999zs} (the constant potential with $\lambda=0$ represents the trivial case, leading to a cosmological constant type of behaviour). The particular case of $\sigma=0$, i.e., of a totally uncoupled setting, was previously studied in, for example, Ref.~\cite{cop}. 

Making use of these variables, we can also define the density parameter and the equation of state parameter for each fluid:
\begin{eqnarray}
\Omega_{\phi}&=&\frac{y^2}{\sqrt{1-x^2}},
\label{omega}\\
\Omega_{m}&=& z^2,
\label{omegac}\\
\Omega_{r}&=&r^2,
\label{omegar}\\
w_{\phi}&=& x^2-1.
\label{w}
\end{eqnarray}

From Eqs.~\eqref{omega}--\eqref{omegar} we rewrite the Friedmann constraint in terms of the dimensionless variables
\begin{equation}
\Omega_{\phi}+ \Omega_{r} + \Omega_{m} =1 \Longrightarrow\frac{y^2}{\sqrt{1-x^2}} + r^2 + z^2=1,
\label{friedcons}
\end{equation}
\noindent which we used to replace $z$ in terms of $x$, $y$, and $r$, reducing the dimension of the dynamical system.

From Eq.~\eqref{Qfrwq} and using the definition of the dynamical variables in \eqref{variables} and Eq.~\eqref{friedcons}, the interaction term can be recast into
\begin{equation}
\frac{2\kappa^2}{3H^3}\, Q =  \sqrt{3}\, \sigma\, y  \left(1-r^2-\frac{y^2}{\sqrt{1-x^2}} \right).
\label{Qdin}
\end{equation}
Taking the expression of the interaction term in Eq.~\eqref{Qdin} and the variables defined in Eq.~\eqref{variables}, we can write the system of autonomous equations, Eqs.~\eqref{kg}--\eqref{fried2}, as
\begin{align}
x'=&(x^2-1) \left[ \frac{\sqrt{3}\, \sigma}{2\, y}  \left(r^2+\frac{y^2}{\sqrt{1-x^2}}-1 \right) \sqrt{1-x^2} \right. \nonumber \\ & \left. \ \ \ \ \ \ \ \ \ \ \ \ \  +3x - \sqrt{3}\, \lambda\, y \right], 
\label{xl} \\
y'=&-\frac{y}{2} \left( \sqrt{3}\, \lambda\, x\, y +2 \frac{H'}{H}\right),
\label{yl} \\ 
r'=&-r \left(2+ \frac{H'}{H}\right) ,
\label{rl}  
\end{align}

\noindent where a prime denotes derivatives with respect to $N \equiv \ln a$ and we have used 
\begin{equation}
\frac{H'}{H}= - \frac{3}{2} \left(1+ \frac{1}{3} r^2  - y^2\, \sqrt{1-x^2} \right).
\label{hlinha}
\end{equation}
\noindent The effective EoS parameter can be identified as the quantity inside the brackets in Eq.~\eqref{hlinha} such that, using the Friedmann constraint, Eq.~\eqref{friedcons},
\begin{equation}
\frac{H'}{H}= - \frac{3}{2} (1+w_{\text{\rm eff}}) \Longrightarrow w_{\text{\rm eff}}=  \frac{1}{3} r^2 - y^2\, \sqrt{1-x^2}.
\label{weff}
\end{equation} 
\noindent Being a global parameter, $w_{\rm eff}$ is the quantity that determines whether the Universe undergoes a period of accelerated ($w_{\rm eff} < -1/3$) or decelerated ($w_{\rm eff} > -1/3$) expansion. For this particular case, we require that 
\begin{equation}
 \frac{1}{3} r^2 - y^2\, \sqrt{1-x^2} < - \frac{1}{3},
\label{wacc}
\end{equation}
\noindent in the present, in order to guarantee an accelerated expansion of the Universe. 

The integration of Eq.~\eqref{weff}, allows us to describe the evolution of the scale factor with time at any given fixed point solution, $\left(x^f,y^f,r^f \right)$, when the dynamical system is in equilibrium,
\begin{equation}
a \propto t^{2/3 \left(1+w^f_{\rm eff} \right)},
\end{equation}
\noindent with $w^f_{\rm eff}\equiv w_{\rm eff} \left(x^f,y^f,r^f \right)$ as defined in Eq.~\eqref{weff}. The limit case in which $w^f_{\rm eff} = -1$ reduces the cosmological evolution to a de Sitter exponential expansion with a constant Hubble rate.

\subsection{\label{sec:ps} Phase space and invariant sets} 

Throughout this study, we rely on the fact that $ 0 \leq \Omega_{\phi} \leq 1$, which translates into a non-negative energy density for the matter and radiation fluids, rendering the allowed region for $x$ and $y$:
\begin{equation}
0\leq x^2 + y^4 \leq1.
\label{rest}
\end{equation} 
This constitutes the physical phase space, i.e., the invariant set which contains all of the orbits with physical meaning.

A glance at the system of Eqs.~\eqref{xl}--\eqref{rl} makes it clear that the dynamical system is invariant under the mappings: $(x,y) \longmapsto (-x,-y)$, $r\longmapsto -r$, and $(x, \lambda, \sigma) \longmapsto (-x,-\lambda,-\sigma)$. This implies that, since there is a full symmetry under simultaneous sign exchanges, we need only to analyse the region of the phase space corresponding to non-negative values of $y$, $r$, and $\lambda$. The negative values would lead to the same dynamics under appropriate reflections. 
Furthermore, we assume $H > 0$ in order to impose an expanding Universe.

\subsection{\label{sec:fp}Fixed points and physical phase space}

The system of autonomous equations \eqref{xl}--\eqref{rl} is fundamentally different for the particular case of $\sigma=0$ and for $\sigma \neq 0$. When $\sigma \neq 0$, there is a singularity present in the system, whenever $y=0$, meaning that the flow is not formally defined over that plane. Therefore, each separate case has a different set of fixed points. 

The fixed points for this system are obtained by setting the left-hand side of the autonomous equations \eqref{xl}--\eqref{rl} equal to zero.
Moreover, we are interested in performing a stability study, and, whenever possible, we do it through linear stability analysis. However, this is not always the case due to the fact that, when $\sigma \neq 0$, there are nonhyperbolic fixed points and this complication will be addressed further on.

The fixed points for the system \eqref{xl}--\eqref{rl} are presented in Table~\ref{table:gamma1}, where, for simplicity, we have defined
\begin{equation}
\begin{aligned}
&y_{\rm D} \equiv \left( \frac{\sqrt{\lambda^4+36}-\lambda^2}{6} \right)^{1/2}\ \ \ \text{and}\\
&y_{\rm S} \equiv \left( \frac{\sqrt{\sigma ^2\, (\sigma -2 \lambda )^2+36}+6}{(\sigma -2 \lambda )^2} \right)^{1/2}.
\end{aligned}
\label{ys}
\end{equation}
Note that Table~\ref{table:gamma1} is divided into three sections, separated by a double horizontal line. The first one corresponds to the fixed points which are only found for the system with $\sigma=0$, the second one contains fixed points which exist independently of the value of $\sigma$, and the last section allows us to present the new fixed point, associated with the interplay between the scalar field and the matter sector when $\sigma \neq 0$.
The corresponding values of the density parameter $\Omega_{\phi}$, Eq.~\eqref{omega}, EoS parameter of the field $w_{\phi}$, Eq.~\eqref{w}, and effective EoS parameter $w_{\rm eff}$, Eq.~\eqref{weff}, for each fixed point are also listed.
Additionally, we present the range of parameters corresponding to an accelerated expansion state, according to Eq.~\eqref{wacc}, and the stability character of each fixed point solution.

For the sake of completeness, before introducing the novelties associated with the system with $\sigma \neq 0$, we briefly present the main features of the previously well-studied system with $\sigma=0$ \cite{cop,Aguirregabiria:2004xd,Landim:2015poa}.

\begin{table*}[t]
\centering
\caption{Fixed points of the system of equations \eqref{xl}--\eqref{rl} considering a dustlike coupled fluid and the corresponding cosmological parameters, as defined in Eqs.~\eqref{omega}, \eqref{w}, \eqref{rest}, and \eqref{weff}. The parameter region corresponding to an accelerated expansion state is evaluated according to Eq.~\eqref{wacc}. The dynamical stability character is inferred by taking into account the eigenvalues of the matrix $\mathcal{M}$ evaluated at each fixed point. The eigenvalues can be found in Appendix~\ref{app:eig}. The form of $y_{\rm D}$ and $y_{\rm S}$ is given in Eq.~\eqref{ys}. The parameter $\widetilde{\lambda}$ is defined in Eq.~\eqref{theta}. Note that, by construction, we are considering $\lambda> 0$.}
\begin{tabular}{ l c c c c c c c c c}
\hline\hline \\[-1.5ex]
Name & $x$ &$y$ & $r$ & $\Omega_{\phi}$ & $w_{\phi}$ & Existence & $w_{\rm eff}$ & Acceleration & Stability \\
\hline \\[-1.5ex]
(O) & $0$ & $0$ & $0$ & $0$ & $-1$ & $\forall \lambda, \sigma=0$ & $0$ & No & Saddle\\
(C) & $0$ & $0$ & $1$ & $0$ & $-1$ & $\forall \lambda, \sigma=0$ & $\frac{1}{3}$ & No & Saddle \\[1ex]
\hline\hline \\[-1.5ex]
(A$_\pm$) & $\pm1$ & $0$ & $0$ & Und. & $0$ & $ \forall \lambda, \sigma$ & $0$ & No & (See text)\\
(B$_\pm$) & $\pm 1$ & $0$ & $1$ & $0$ & $0$ & $ \forall \lambda, \sigma $ & $\frac{1}{3}$ & No & Repeller\\
(D) & $\frac{\lambda y_{\rm D}}{\sqrt{3}} $ & $y_{\rm D}$ & $0$ & $1$ & $\frac{\lambda^2 y_{\rm D}^2}{3} -1$ & $ \forall \lambda, \sigma $ & $\frac{\lambda^2 y_{\rm D}^2}{3}-1$ & $\lambda^4<12$ & \begin{tabular}{@{}c@{}}Stable for $\sigma \geq \widetilde{\lambda}$ and \\ saddle otherwise \end{tabular} \\[1ex]
\hline\hline \\[-1.5ex]
(S) & $\frac{2 \sqrt{3}}{(2 \lambda -\sigma ) y_{\rm S}}$ & $y_{\rm S}$ & $0$ &$\frac{(2 \lambda-\sigma) y_{\rm S}^3}{\sqrt{(2\lambda-\sigma)^2 y_{\rm S}^2-12}}$ & $\frac{12}{(2\lambda-\sigma)^2 y_{\rm S}^2}-1$ & $\sigma\leq \widetilde{\lambda}$ & $\frac{\sigma}{(-2 \lambda + \sigma)}$ & $\ \sigma \leq \widetilde{\lambda} \wedge \sigma < - \lambda\ $ & Stable for $\sigma \leq \widetilde{\lambda}$ \\[1ex]
\hline
\end{tabular}
\label{table:gamma1}
\end{table*}

\subsubsection{Case of $\sigma=0$} \label{sec:s0}

The fixed points for the system of equations \eqref{xl}--\eqref{rl} with $\sigma=0$ are listed in the first and second sections of Table~\ref{table:gamma1}.
In this case, all of the fixed points are hyperbolic, and, therefore, their stability character can be inferred simply through linear stability analysis.
Some comments can be stated regarding the existence and stability of the fixed points:

\begin{itemize}
\item The \textit{matter dominated} fixed point (O) represents a matter dominated solution ($\Omega_\phi=0$, $\Omega_r=0$), which is consistent with a null effective equation of state, $w_{\rm eff } = w_{m}=0$. The scalar field is negligible near this fixed point. It will always be a saddle point, which attracts orbits along the $x$ axis and repels them towards the $y$ axis. 

\item The \textit{kinetic with matter} fixed points (A$_\pm$), due to the indetermination in Eq.~\eqref{omega}, can represent matter dominated solutions or scalar field dominated solutions, depending on the direction of the trajectories approaching them: for trajectories close to the boundary, $x^2+y^4=1$, they represent scalar field dominated solutions, whereas for trajectories close to the boundary $y=0$, they stand for matter dominated solutions. Accordingly, $w_{\rm eff}=w_{m}= w_{\phi}=0$ and these fixed points are not capable of providing accelerated expansion. They are always saddle points of the system.

\item Points (B$_\pm$) stand for \textit{kinetic with radiation domination} solutions, characterised by $\Omega_{r} = r^2 = 1$, and, consistently, a constant effective equation of state $w_{\rm eff} = w_{r} = 1/3$. Also, we find $w_\phi = 0$, meaning that near these points the scalar field acts as a dustlike fluid. Moreover, they are independent of the choice of the parameters, and we find that they are always repellers since all of the eigenvalues are realvalued and positive. Being the only repellers makes them the only possible past attractors of the system.

\item The \textit{radiation dominated} fixed point (C) is characterised by $w_{\rm eff} = 1/3$ and $w_\phi = -1$, which means that the scalar field (whose contribution, in practical terms, is negligible) freezes and is akin to a cosmological constant. It is always present in the phase space and is a saddle point. This point is the equivalent on the $r \equiv 0$ plane to the point (O) on the $r \equiv 1$ plane.

\item The fixed point (D) corresponds to a \textit{scalar field dominated} solution since $\Omega_\phi = 1$ for every parameter value. This fixed point has an explicit dependence on $\lambda$, it exists for every constant value of the parameters, and it always lies on the boundary $x^2+y^4=1$. We find that $w_\phi=w_{\rm eff}=\frac{\lambda^2 y_{\rm D}^2}{3}-1$. Whenever $\lambda^4<12$, it lies inside the area of the phase space where the Universe undergoes accelerated expansion \cite{Gumjudpai:2005ry}. It is the global attractor of the system. Moreover, in the limit $\lambda \rightarrow 0$, it features a cosmological constant type of behaviour.
\end{itemize} 

\begin{figure}[b]
  \includegraphics[width=0.8\linewidth]{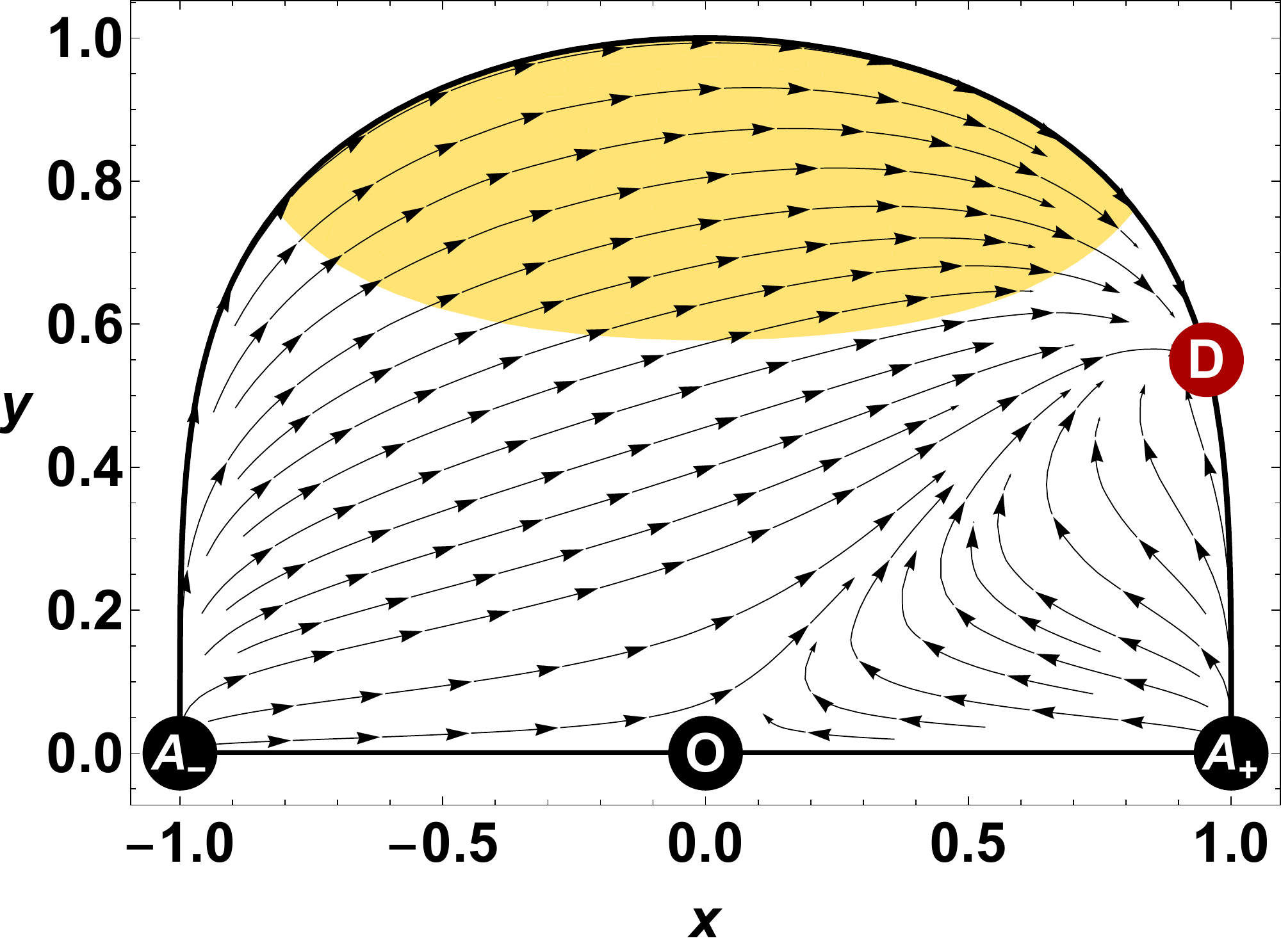}
  \caption{\label{fig:phase0} Two-dimensional reduced phase portrait ($ r \equiv 0$) of the dynamical system given by Eqs.~\eqref{xl} and \eqref{yl}, for $\lambda=3$ and $\sigma=0$. The yellow shaded region denotes the parameter region for which the Universe undergoes accelerated expansion, Eq.~\eqref{wacc}. The attractor is the fixed point (D). The orbit connecting the fixed points (O) and (D) clearly divides the phase space into two invariant sets.}
\end{figure}

The dynamical analysis of this setting is straightforward. Since the evolution of the (uncoupled) $r$ component is trivial, we focus only on the reduced phase space in the $x$-$y$ plane, which is depicted in Fig.~\ref{fig:phase0} for $\lambda=3$. The future attractor is always the fixed point (D), whose presence inside the yellow shaded region, depicting accelerated expansion, is merely a function of $\lambda$. All of the orbits of the full phase space are solutions which connect the fixed points (B$_\pm$) or (A$_\pm$) to (D), except for the ones connecting (B$_\pm$)
 to (A$_\pm$), (B$_\pm$) or (A$_\pm$) to (O) or (C), (C) to (O), and (C) or (O) to (D). 

This model can be applied in order to describe the transition between the matter and dark energy dominated eras. However, in this framework, the Universe will inevitably become totally dark energy dominated and expand indefinitely. Accordingly, the fixed point solution can reproduce the present observed matter content of the Universe. Therefore, in order to successfully reproduce the observed matter content of the Universe, one has to select specific conditions in such a way that the attractor is not yet reached today, and we are living in a transient period of the Universe, already with accelerated expansion. This specific set of conditions leads to the fine-tuning problem.

\subsubsection{Case of $\sigma \neq 0$}

The fixed points for the system of equations \eqref{xl}--\eqref{rl} with $\sigma \neq 0$ are listed in the second and third sections of Table~\ref{table:gamma1}.
As stated before, the dynamical system defined in Eqs.~\eqref{xl}--\eqref{rl} with $\sigma \neq 0$ is singular for $y=0$. Away from the singularity, it is always possible to use the linear approximation in order to infer the stability character of the fixed points. For the points which lay on the $y=0$ plane, (A$_\pm$) and (B$_\pm$), more care must be taken. 

\begin{itemize}

\item The kinetic with radiation domination points (B$_\pm$) are easily regularised because the condition $r=1$ fully specifies the asymptotic behaviour at the singular point. Their stability nature can then be studied through the same linear stability techniques. The previous analysis still holds, as they are still the only repellers of the system. 

\item For the case of the kinetic with matter fixed points (A$_\pm$), we find that the indeterminacy can be lifted only for one of them, (A$_+$) or (A$_-$), depending on the sign of $\sigma$. This point is found to be a saddle, and a repeller on the $r \equiv 0$ plane, as in the case $\sigma = 0$. The behaviour of the flow near the other point has to be studied numerically and will be explained in detail in Sec. \ref{sec:amm}.

\item The stability character of the scalar field dominated fixed point (D), the attractor of the uncoupled case, may be altered for nonzero values of $\sigma$. We find that for a fixed value of $\lambda$, a transcritical bifurcation involving (D) and (S) can occur, changing (D)'s stability nature (see Appendix~\ref{sec:mybif} for more details). Since, for symmetry reasons, we are considering only the case where $\lambda >0$, the fixed point (D) is stable in the parameter region defined as
\begin{equation}
\sigma \geq \widetilde{\lambda},
\label{stablee}
\end{equation}
\noindent with 
\begin{equation}
\widetilde{\lambda} \equiv \lambda \left(1 -\sqrt{1+36/\lambda ^4} \right).
\label{theta}
\end{equation}
\noindent Otherwise it is a saddle point.
This fixed point solution can feature accelerated expansion independently of the value of the coupling parameter $\sigma$. 

\item The new fixed point (S) corresponds to a \textit{conformal scaling} fixed point, as the cosmological parameters $\Omega_\phi$, $w_\phi$, and $w_{\rm eff}$ depend on the values of $\lambda$ and $\sigma$. %This means that, near the fixed point solution, the dynamics of the scalar field naturally self-adjusts to that of matter and the corresponding energy densities become proportionally constant. 
Owing to the restriction in Eq.~\eqref{rest} and $\lambda>0$, this point is present in the phase space only when
\begin{equation}
\sigma \leq \widetilde{\lambda},
\label{l1} 
\end{equation}
\noindent with $\widetilde{\lambda}$ as defined in Eq.~\eqref{theta}. It can provide accelerated expansion whenever $\sigma < - \lambda$, and it is stable for Eq.~\eqref{l1}. This means that, when its cosmological existence is verified, (S) is the attractor of the system.
\end{itemize}

By looking at the value of $w_{\rm eff}$ for the fixed point (S) in Table~\ref{table:gamma1}, one interesting feature is that, for a fixed value of the parameters, we can achieve an accelerating Universe by means of this scaling solution. This is of great physical interest since the scaling solution allows for a more natural description of the comparable values in the scalar field and matter energy densities, alleviating the cosmic coincidence problem, as an everlasting expanding solution with $\Omega_\phi \approx 0.7$ can now be achieved, independently of the set of initial conditions. 

\begin{figure}[t]
  \includegraphics[width=0.9\linewidth]{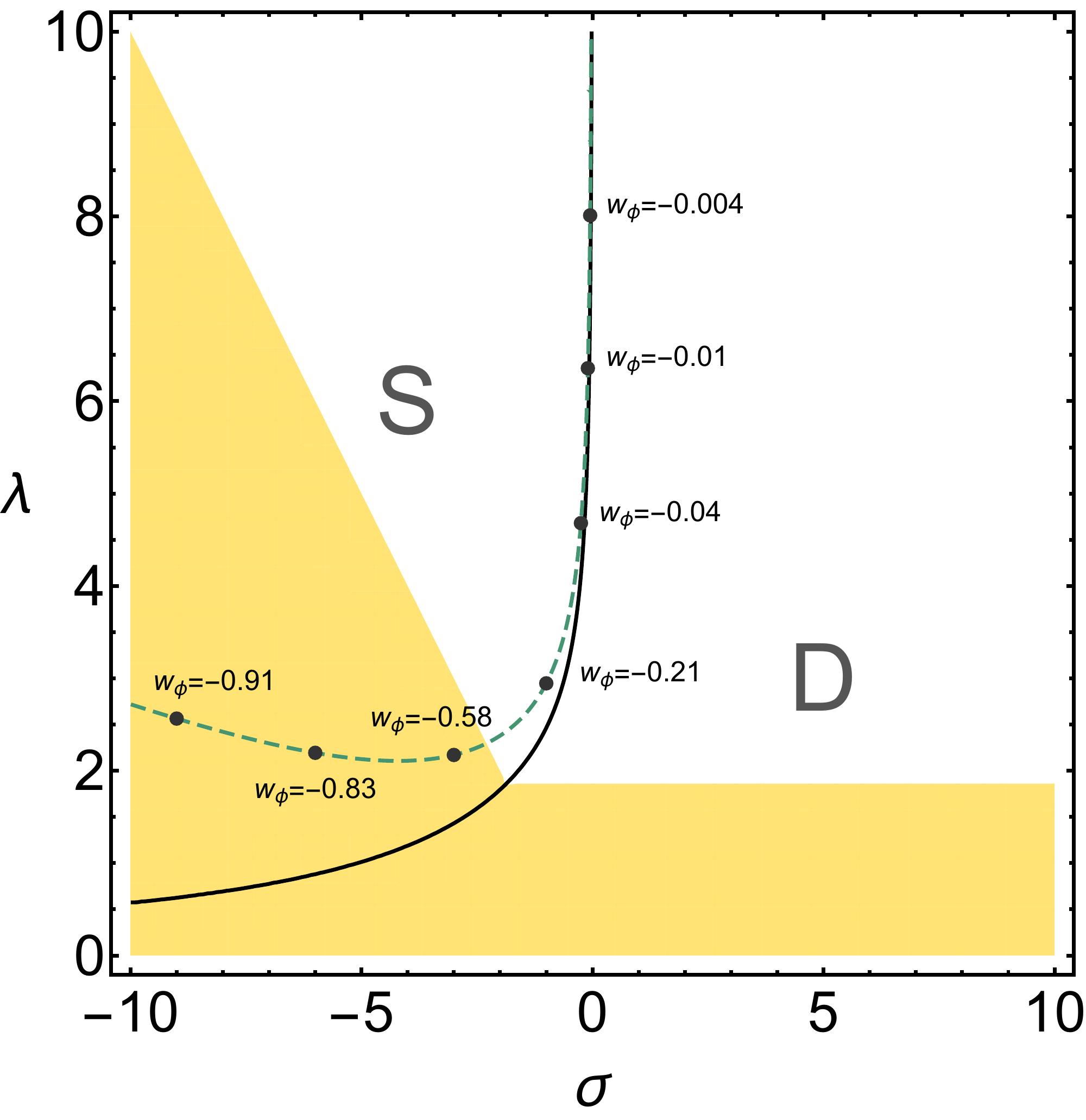}
  \caption{\label{fig:regs} Illustration of the attractor of the system, according to the ($\sigma$, $\lambda$) parameter region, as defined in Eqs.~\eqref{stablee} and \eqref{l1}. The regions are separated by a solid black line and are labelled by the critical point that is stable in that region (according to Table~\ref{table:gamma1}). The shaded (yellow) areas illustrate the parameter values for which each attractor is accelerated. The dashed curve indicates the parameter values where $\Omega_\phi (\text{S}) = 0.7$, with some examples of the corresponding values of $w_\phi (\text{S})$ labelled. }
\end{figure}

In Fig.~\ref{fig:regs}, we present a depiction of the attractor of the system according to the parameter region. Note that, for every value of the parameters $\lambda$ and $\sigma$, there is always one -- and only one -- possible attractor of the system.

\begin{figure*}[t]
   \subfloat[$\sigma=0.5$]{\includegraphics[width=0.33\linewidth]{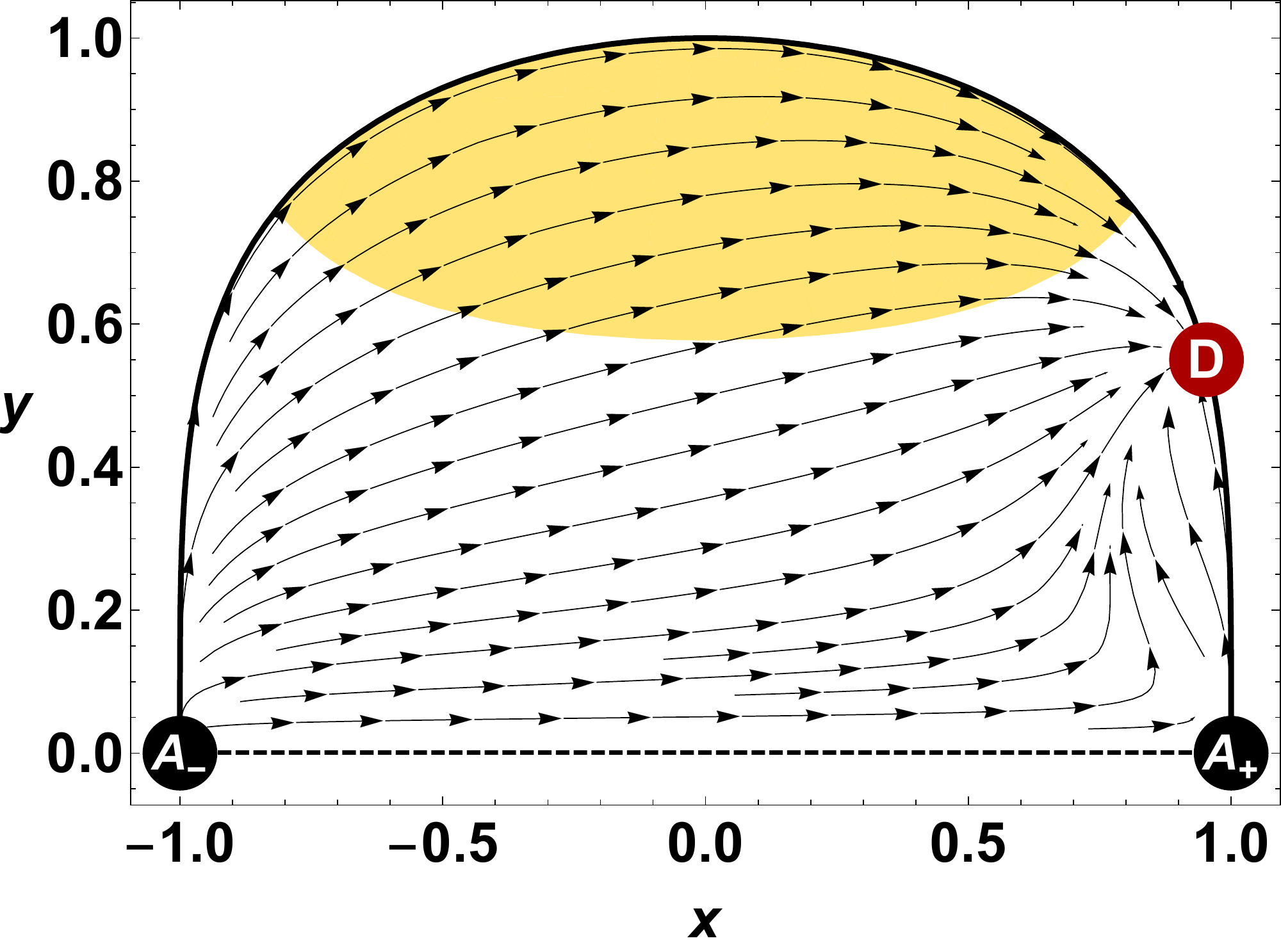}} 
        \hfill
    \subfloat[$\sigma =-0.5$]{\includegraphics[width=0.33\linewidth]{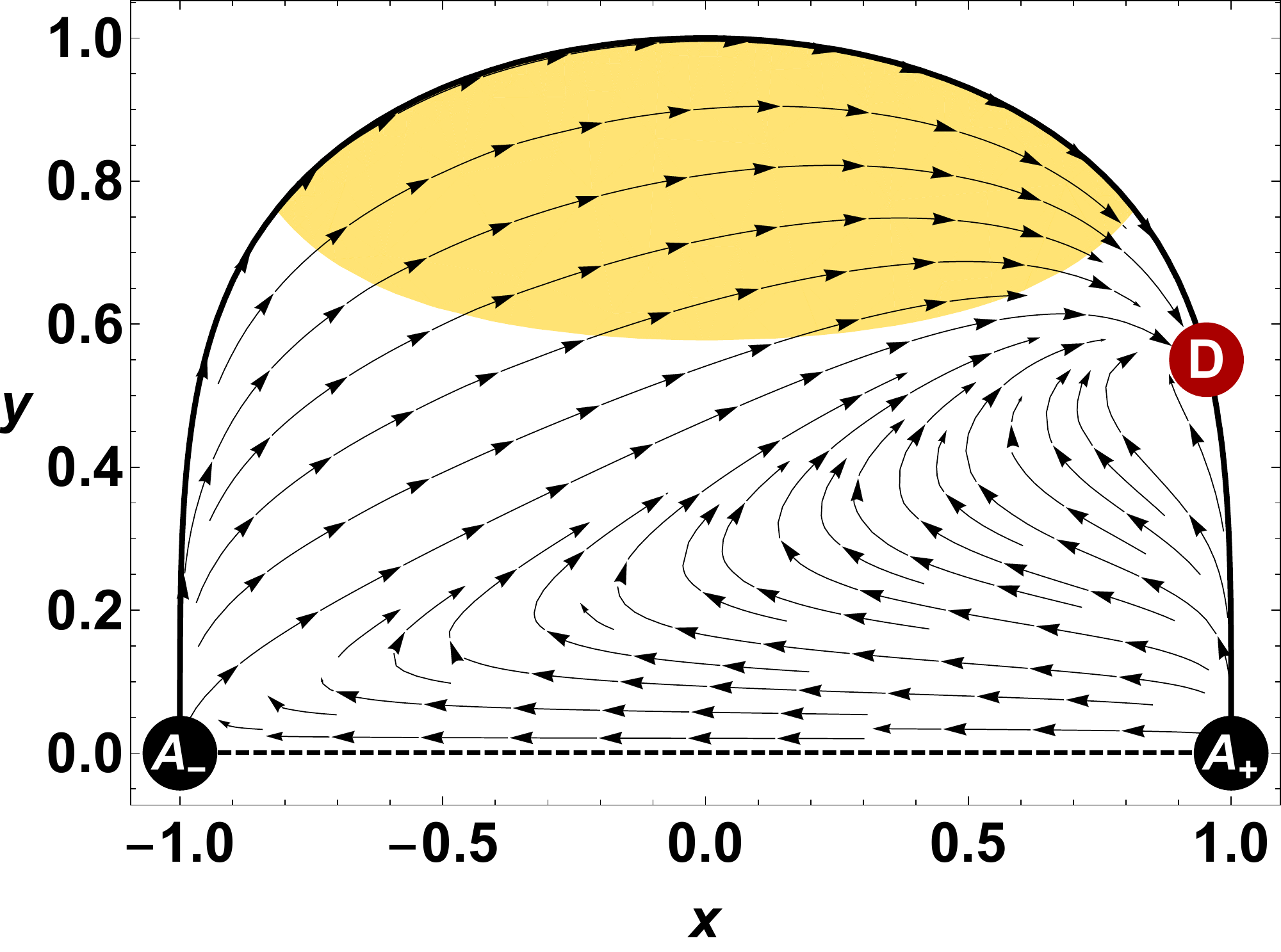}}
   \hfill
    \subfloat[$\sigma =\widetilde{\lambda}$]{\includegraphics[width=0.33\linewidth]{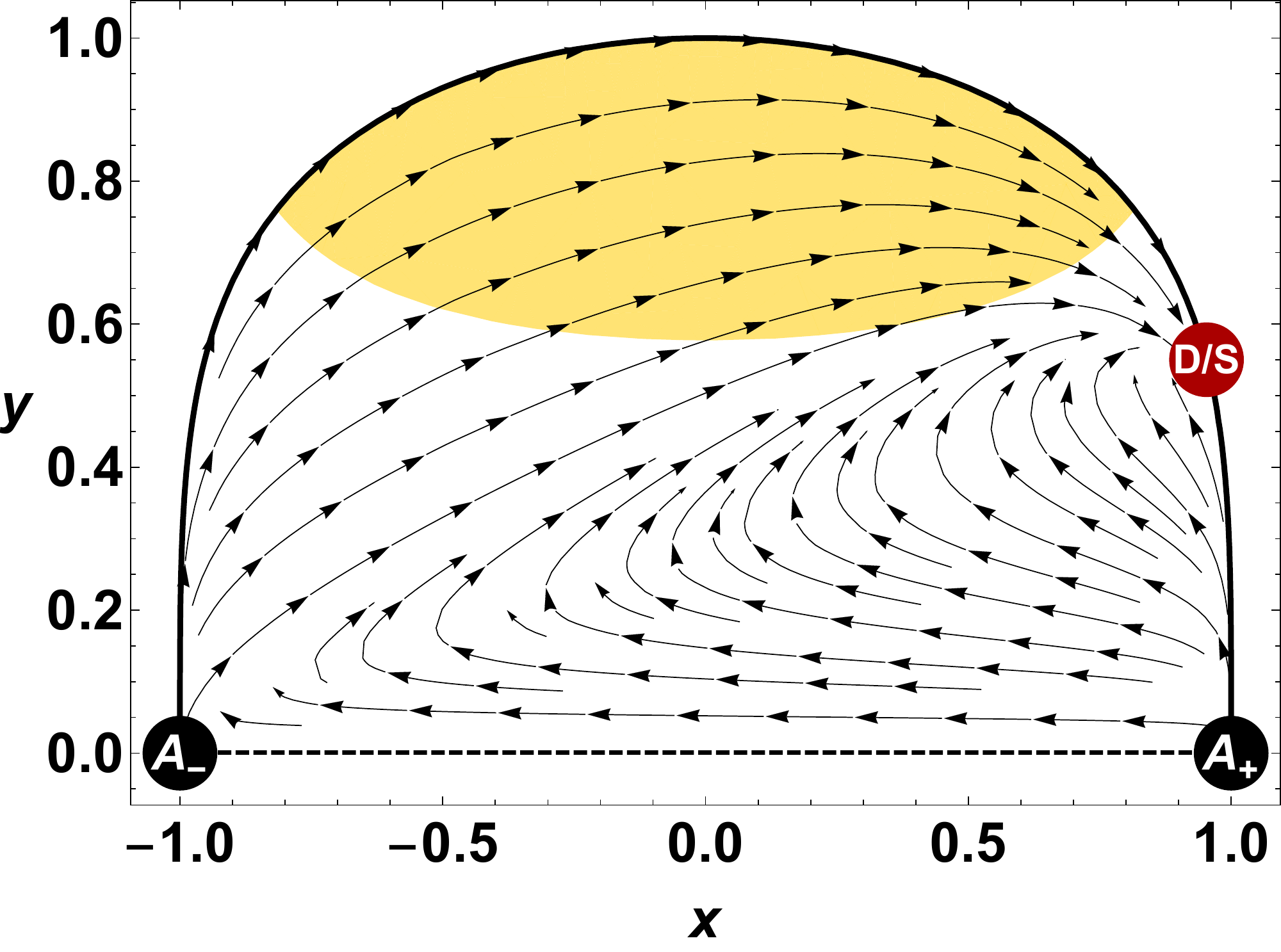}}
        \hfill
    \subfloat[$\sigma=-3$]{\includegraphics[width=0.33\linewidth]{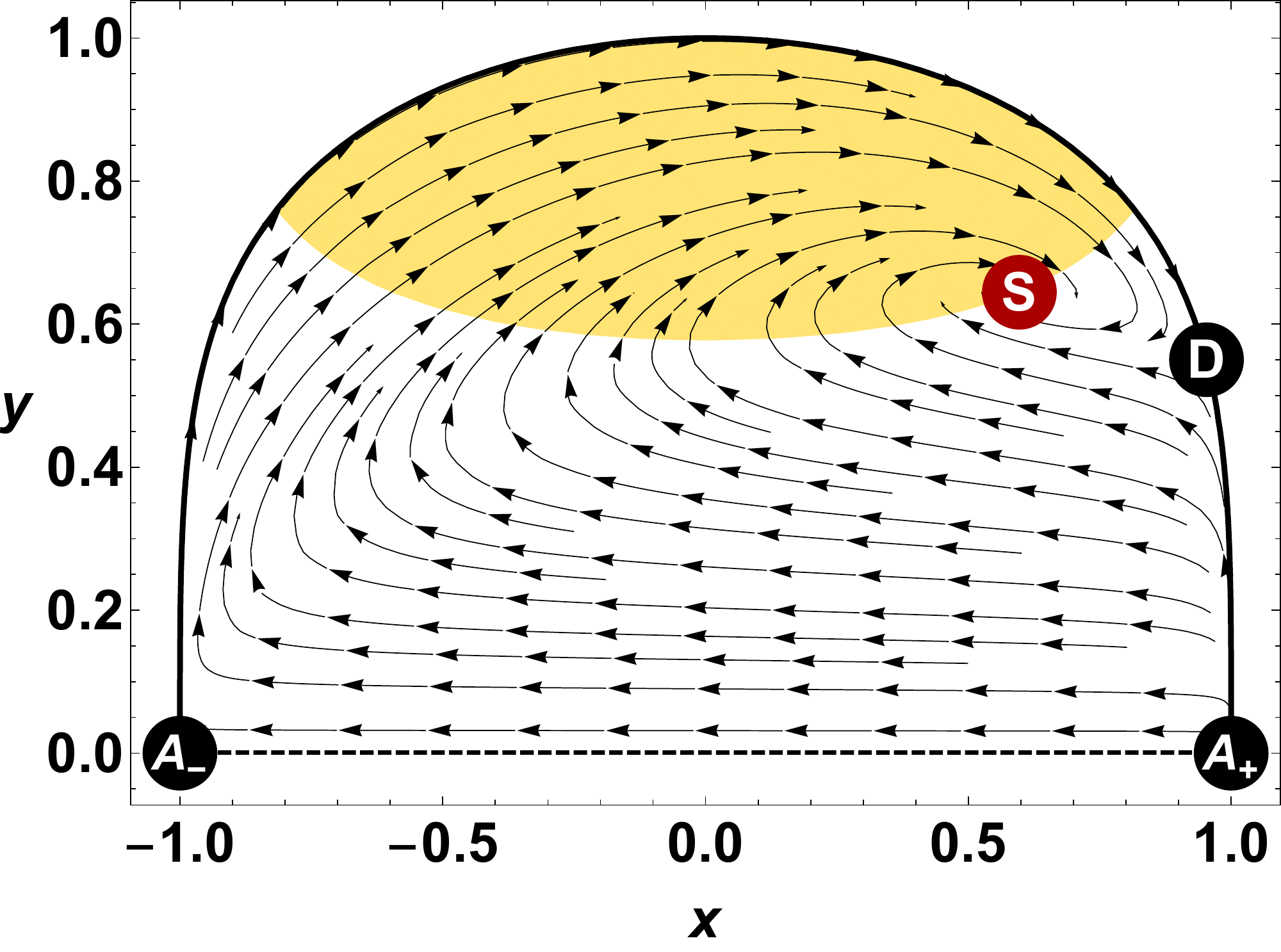}}
        \hfill
    \subfloat[$\sigma=-10$]{\includegraphics[width=0.33\linewidth]{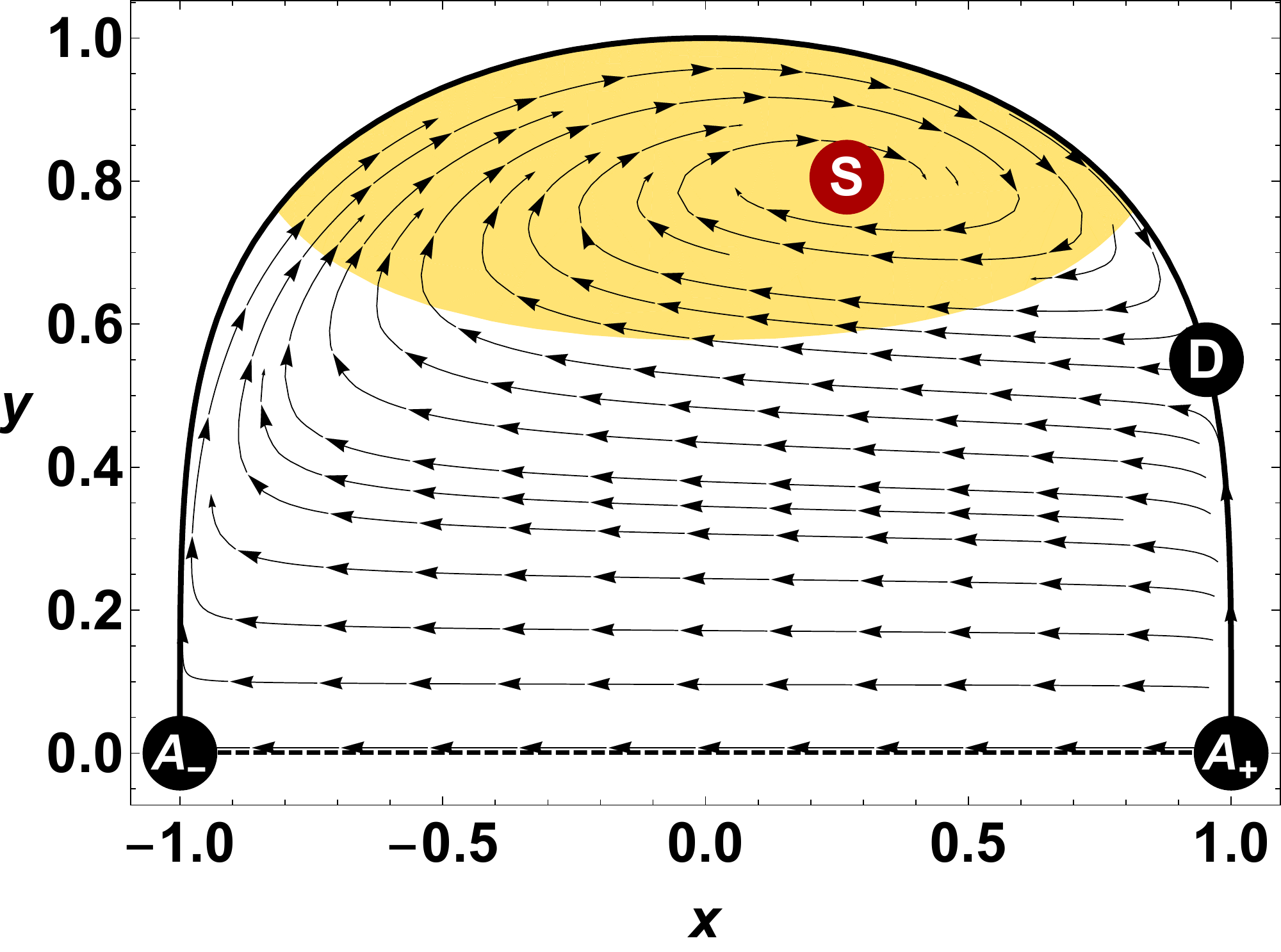}}
        \hfill
    \subfloat[$\sigma =-50$]{\includegraphics[width=0.33\linewidth]{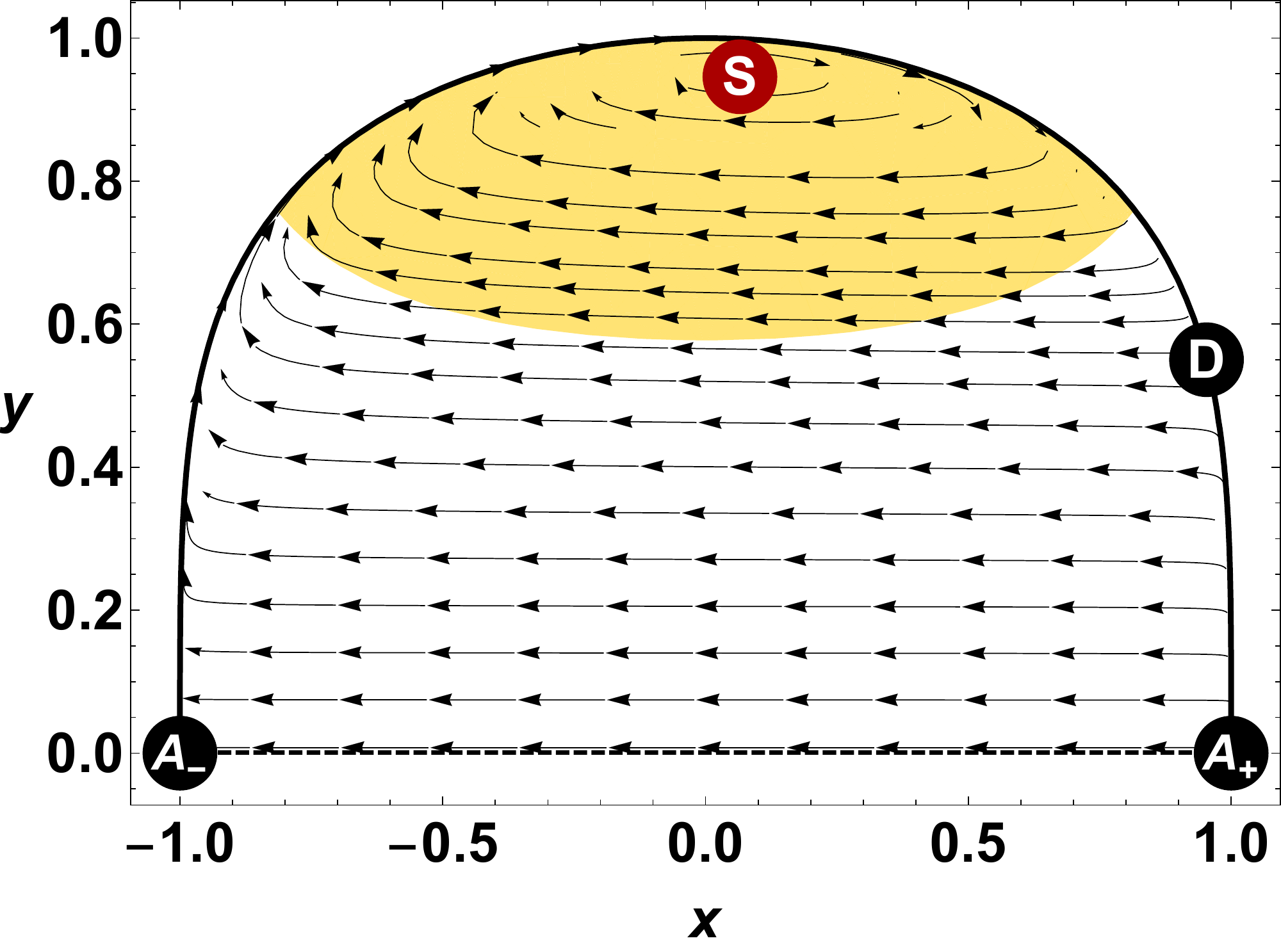}}
  \caption[Phase portrait projections for the uncoupled and coupled tachyonic models]{\label{fig:ps2d} Two-dimensional reduced phase portraits of the dynamical system given by Eqs.~\eqref{xl} and \eqref{yl} (see text), for $r \equiv 0$, $\lambda=3$ and different values of the coupling parameter $\sigma$. The black and red points correspond to the fixed points defined in Table~\ref{table:gamma1}. The yellow/shaded region corresponds to the region where the Universe undergoes accelerated expansion, as defined in Eq.~\eqref{wacc}, and $\widetilde{\lambda}$ is defined in Eq.~\eqref{theta}.}
\end{figure*}

In Fig.~\ref{fig:ps2d}, we plot the phase space for $\lambda=3$ and different values of $\sigma$, for $r \equiv 0$, since the dynamics associated with the $r$ component is trivial. This means that only the late-time matter to the dark energy transition is illustrated, where the contribution of the relativistic fluids can be neglected. All of the orbits correspond to solutions connecting the fixed points (B$_\pm$) or (A$_\pm$) to either (D) or, when it exists, (S). The only exceptions are the orbits connecting the fixed points (B$_\pm$) to (A$_\pm$) (with $x \equiv \pm 1$ and $y \equiv 0$) and the orbit from (D) to (S). 

For $\sigma > \widetilde{\lambda}$, there are only three fixed points in the $r \equiv 0$ plane, and the dynamics is as portrayed in Figs.~\ref{fig:ps2d} (a) and \ref{fig:ps2d} (b).
%In Figure \ref{fig:ps2d} (b), a qualitatively depiction of the dynamics in this parameter region is presented for $\lambda=3$ and $\tau=-10$.

For $\sigma < \widetilde{\lambda}$, there are four fixed points present in the phase space. Qualitatively, (D) turns into a saddle, and a new attractor (S) arises. Depending on the value of $\sigma$, the point (S) can be located anywhere starting from the fixed point (D), along an arc of the parabola $(x,y,r) = (x_{\rm S},y_{\rm S},0)$, towards the point $(0, 1, 0)$, as $\sigma \rightarrow - \infty$. This feature is depicted in Fig.~\ref{fig:ps2d} (d) -- Fig.~\ref{fig:ps2d} (f).
The fixed point (S) is found only inside the accelerated expansion region for $ \sigma < - \lambda$.
Even when the attractor (S) itself does not meet the previous condition, there are many solutions which feature a transient period of accelerated expansion by crossing the yellow shaded region.

In the limit case where $\sigma = \widetilde{\lambda}$, the fixed points (D) and (S) coincide and become the attractor of the system. This limit case is portrayed in Fig.~\ref{fig:ps2d} (c).

Note that, for $\lambda>0$ and according to Eq.~\eqref{l1}, near the attractor, (D) or (S), it is always true that $\dot{\phi} > 0$ (see Table~\ref{table:gamma1}). Additionally, the sign of the coupling $Q$, as defined in Eq.~\eqref{Qdin}, is the same as the sign of $\sigma$. This defines the direction of energy exchange in Eqs.~\eqref{fluidcon} and \eqref{fieldcon}: if $\sigma>0$, the matter fluid is granting energy to the dark energy field and, on the contrary, if $\sigma<0$, it is the DE field which sources the matter sector. This means that, whenever the scaling solution is allowed to exist, near the attractor, energy is always being transferred from the $\phi$ field to the matter component.

\subsection{Dynamics of (A$_\pm$)} \label{sec:amm}

As discussed before, the points (A$_\pm$) are located on the boundary of the phase space, where the flow is not formally well defined. Indeed, from a mathematical point of view, the system of Eqs.~\eqref{xl}--\eqref{rl} is singular for $\sigma \neq 0$ and $y=0$. 

\begin{figure}[t]
  \includegraphics[width=0.8\linewidth]{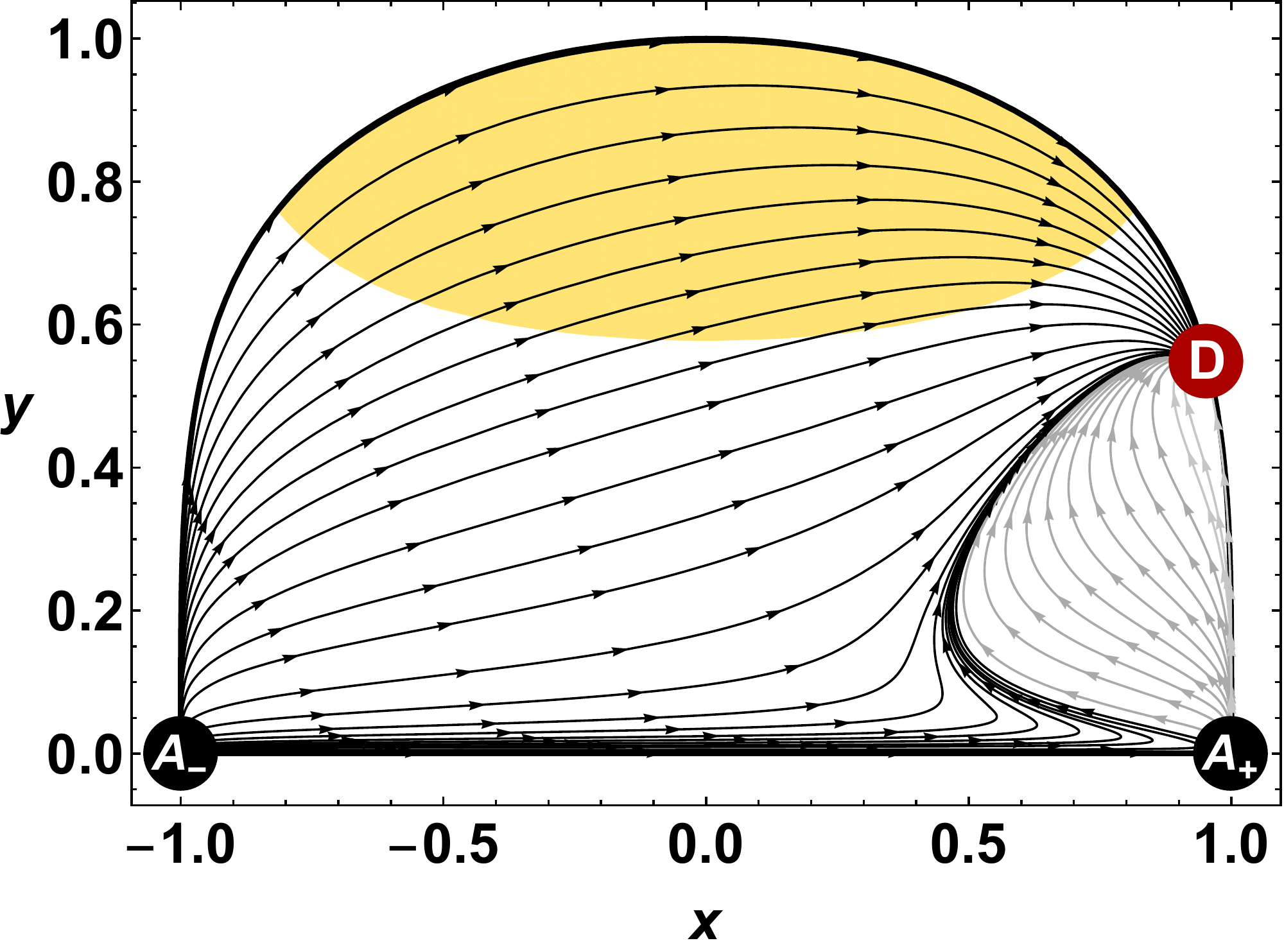}
  \caption{\label{fig:phase} Two-dimensional reduced phase portrait of the dynamical system given by Eqs.~\eqref{xl} and \eqref{yl} (see the text), for $r \equiv 0$, $\lambda=3$, and $\sigma=0.1$. The orbits which have (A$_-$) and (A$_+$) as past attractors are shown in black and gray, respectively. There is clearly a separatrix dividing the two sectors, whose relative size depends on the value of $\sigma$. This is a reminiscent effect of the separatrix given by the orbit connecting the fixed points (O) and (D) for the case $\sigma=0$.}
\end{figure}

Let us consider $\sigma > 0$. The following treatment is analogous for $\sigma \leq 0$, with (A$_+$) $\leftrightarrow$ (A$_-$). Also, let us take $r \equiv 0$ (as represented in Fig.~\ref{fig:phase}), i.e., the dynamics in the $x$-$y$ plane, since the dynamics associated with the $r$ component is trivial. Note that, for $\sigma=0$, in the $x$-$y$ plane, the points (A$_\pm$) are repelling nodes, even though they are saddle points of the overall system, given their attracting character over the $r$ component. As we have seen before, the orbit connecting the fixed points (O) and (D) divides the phase space into two invariant sets with past attractors (A$_-$) or (A$_+$) (see Fig.~\ref{fig:phase0}).

For any $\sigma \neq 0$, we find that the fixed point (O) disappears and (A$_+$) becomes a degenerate fixed point, as if (O) had migrated to the position of (A$_+$), generating a fixed point that combines the two. As a result of this collapse, there is a saddle sector and a repelling node sector associated with the fixed point (A$_+$).
The boundary between the two sectors is found only numerically and is illustrated in Fig.~\ref{fig:phase} for $\sigma=0.1$.
The relative size of the sectors is controlled by the value of $\sigma$. For $\sigma$ close to zero, there is a reminiscent effect of the presence of the fixed point (O), and the node sector is dominant (see Fig.~\ref{fig:phase}). This scenario is very similar to the $\sigma=0$ case, where the phase space is divided into two invariant sets, each with (A$_+$) or (A$_-$) as past attractors. Therefore, even though the limit $\sigma \rightarrow 0$ is noncontinuous, it corresponds to a smooth transition for almost all of the initial conditions. For increasing values of $\sigma$, this effect fades and the saddle sector becomes progressively dominant. In the limit where $\sigma \gg 1$, the node sector becomes negligible and (A$_+$) presents an effective saddlelike behaviour [see Fig.~\ref{fig:ps2d} (a) and \ref{fig:ps2d} (b)].

Therefore, albeit mathematically singular, the limit $\sigma \rightarrow 0$ is physically regular. In cosmological terms, there is still the same possible ``choice" of past attractors for different initial conditions, meaning that there are still a certain set of orbits which have (A$_+$)/(A$_-$) as their past attractors, as was the case for $\sigma=0$. However, the relative size of this set depends on the value of $\sigma$. 

\section{Effective Potential} \label{sec:eff}

An immediate consequence of the energy exchange between the fluids is that an interaction term is present in the expression for the energy density of the pressureless coupled fluid. This can be derived by integration of the continuity equation for the matter fluid, Eq.~\eqref{fluidcon}, with the coupling $Q$ term given by Eq.~\eqref{Qfrwq}, leading to
\begin{equation}
\rho_{m}=\rho_{m_{,0}} \left( \frac{a}{a_0} \right)^{-3} \left( \frac{\phi}{\phi_0} \right)^{\theta},
\label{rhocn}
\end{equation}
\noindent where $\rho_{m_{,0}}$ and $\phi_0$ are constants, $a_0 \equiv 1$ is the value of the scale factor today, $\theta \equiv - \sigma/ \lambda$ and the field $\phi$ can be expressed in terms of the potential defined in Eq.~\eqref{cons}. This can be studied, according to the present observational constraints, in order to restrict the possible values of the parameters $\lambda$ and $\sigma$.
It is also possible to define an effective potential associated with the scalar field by rewriting the coupled equation of motion, Eq.~\eqref{kg}, as:
\begin{equation}
\ddot{\phi} + \left( 1-\dot{\phi}^2 \right) \left( 3H \dot{\phi} + \frac{V_{, \phi}^{\rm eff}}{V^{\rm eff}} \right) =0,
\label{pef}
\end{equation}
\noindent where, for a pressureless matter sector,
\begin{equation}
\frac{V_{, \phi}^{\rm eff}}{V^{\rm eff}} = \frac{V_{, \phi}}{V} + \frac{\sqrt{1-\dot{\phi}^2}}{V}\, \frac{C_{, \phi}}{2C}\, \rho_{m},
\end{equation}
\noindent with $\rho_{m}$ given in Eq.~\eqref{rhocn}. By doing so, equation \eqref{pef} resembles the uncoupled equation of motion. Taking the inverse square potential and a power-law conformal coupling function, as described in Eq.~\eqref{cons}, the collective effect of both the potential associated with the scalar field and the coupling is described by an effective potential of the form
\begin{equation}
V^{\rm eff} (\phi)=V(\phi) \cdot Z(\phi),
\label{Vef2}
\end{equation}
\noindent where $V(\phi)$ is the inverse square potential defined in Eq.~\eqref{cons} and we have defined
\begin{equation}
Z(\phi) \equiv  \exp \left[ \frac{\sqrt{1-\dot{\phi}^2}}{ V_0^2}\, \frac{\rho_{m_{,0}} }{ a^{3}}\, \frac{\theta}{\theta+2}\, \frac{ \phi^{2+\theta } }{\phi_{0}^{\theta}} \right].
\label{zdef}
\end{equation}
\noindent where $V_0^{\rm eff} = V_0^2 / \phi_0^2$ is a constant of integration with units of (mass)$^4$. With this definition, the first term on the rhs of Eq.~\eqref{Vef2} corresponds to the standard inverse square potential, whereas the second term accounts for the effect of the coupling. This means that, when coupled to the matter sector, the evolution of the $\phi$ field is driven by the properties of the effective potential, which becomes ``matter dependent''. Accordingly, the direction of energy exchange depends upon the evolution of the field and the combined sign of $\sigma$ and $\lambda$.

\begin{figure*}[t]
    \subfloat{\includegraphics[width=0.47\linewidth]{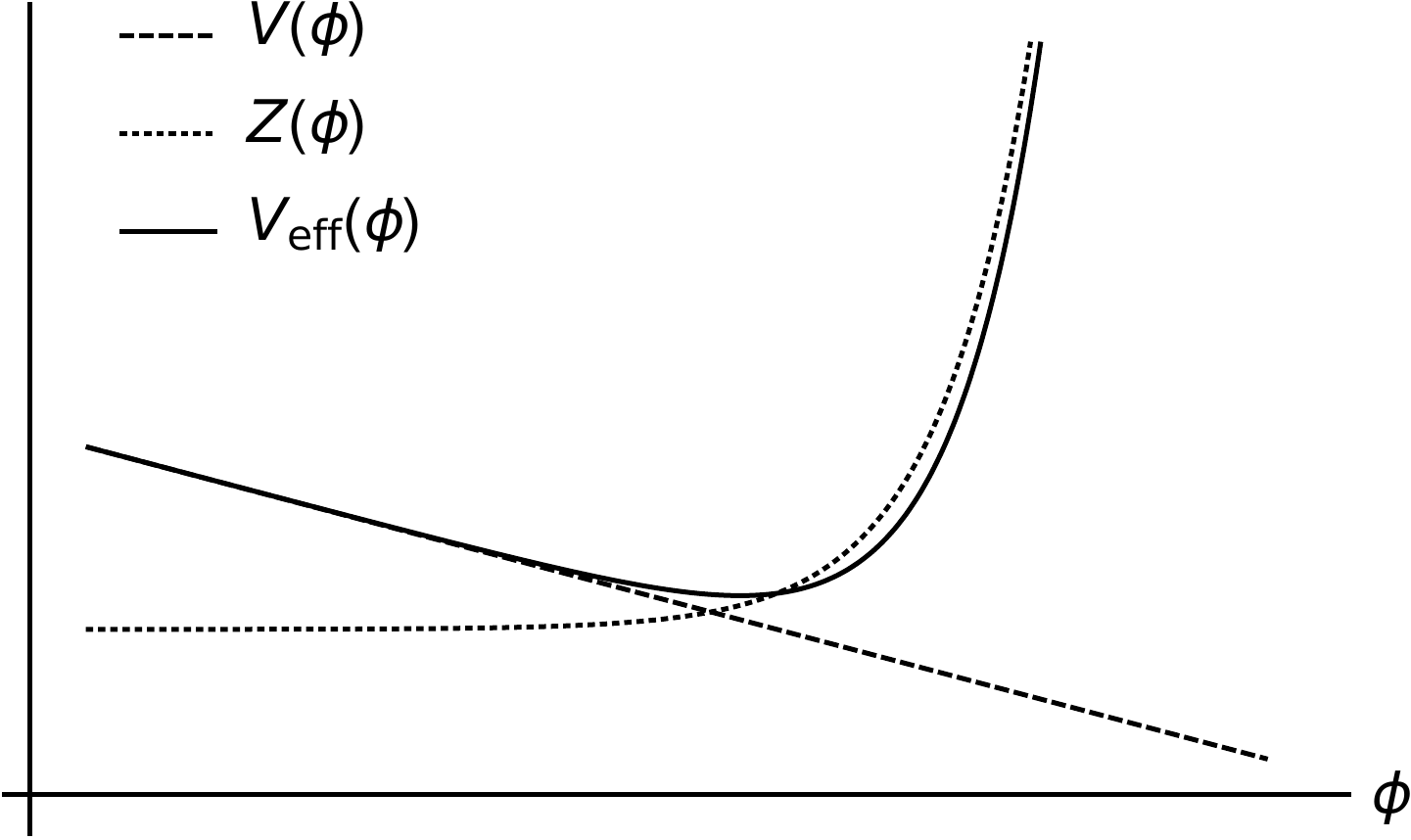}} 
    \hfill
    \subfloat{\includegraphics[width=0.47\linewidth]{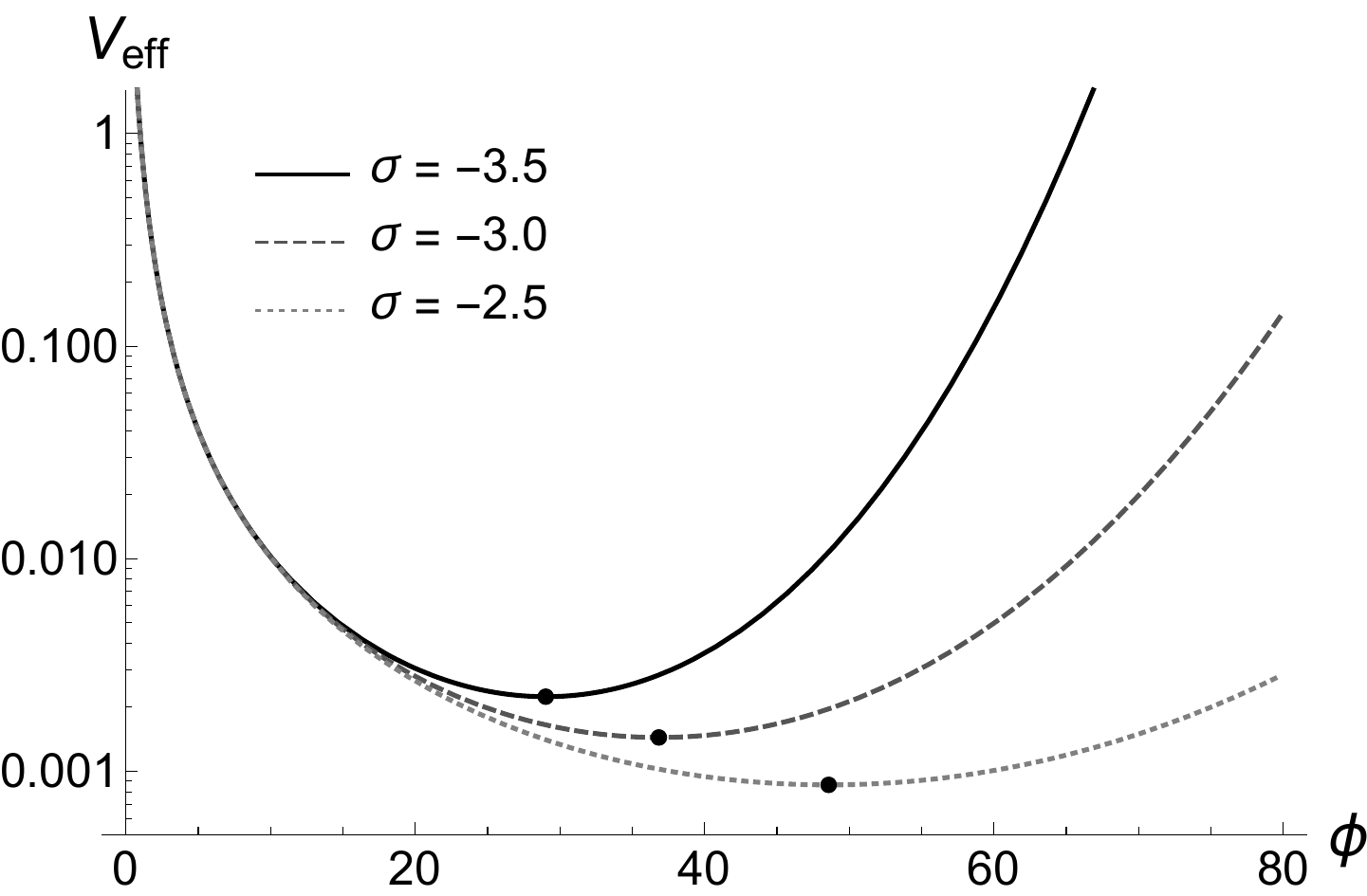}}
  \caption[Effective potential for the conformally coupled tachyonic model]{(Left panel) Illustration of the composition of the effective potential from the combined effect of the inverse square potential and the coupling, as described in Eq.~\eqref{Vef2}. This qualitative representation is made in logarithmic scales for the specific case where $n=1$. (Right panel) Illustration of the effective potential as described in Eqs.~\eqref{Vef2} and \eqref{zdef} for different values of $\sigma$ and constant value of $\lambda=3$ and $B =5 \times 10^4 $ as defined in Eq.~\eqref{bdef}. Decreasing the absolute value of $\sigma$ has the effect of shifting the position of the minimum to larger values of $\phi$ (in absolute value). The minimum point for each curve is marked with a black dot.}
  \label{effpot}
\end{figure*}
From Eq.~\eqref{zdef}, it is clear that, taking $\sigma \rightarrow 0$, which corresponds to the limit in which the coupling term vanishes, we recover the inverse square potential $V(\phi)$. In Fig.~\ref{effpot} (left panel), we present an illustration of the composition of the effective potential, constructed from the combination of the inverse square potential and the conformal coupling function.

Motivated by the study of the parameter space with physical interest performed in the previous sections, we take $\lambda>0$, $\sigma<0$, and $\rho_{m_{,0}} \geq 0$. In this case, it is easy to verify that $V^{\rm eff}$ is a function with one minimum value corresponding to 
\begin{equation}
 \phi_{m} \simeq  \left[   \frac{B}{\theta} \right]^{\frac{1}{2 +\theta}}, 
 \label{phim}
\end{equation}
 \noindent with 
\begin{equation} 
 B  \equiv 2\, V_0^2\,  a^{3}\,   \frac{\phi_{0}^{\theta}}{\rho_{m_{,0}}} 
 \label{bdef}
\end{equation}
\noindent and $\phi_{m} \neq 0$. Note that we have implicitly assumed that $\sqrt{1-\dot{\phi}^2} \sim \mathcal{O} (1)$ as the field approaches the minimum of the potential. 
 
 The position of the minimum depends on the scale factor, $a$, and on the parameters $\lambda$, through $V_0$, and $\sigma$. All of the other variables are considered fixed.

An illustration of the shape of the effective potential for different values of $\sigma$ and $B=5 \times 10^4$ is presented in Fig.~\ref{effpot} (right panel). It is easy to see that, by decreasing the absolute value of $\sigma$ the minimum of the potential is shifted to larger values of $\phi$. When $\sigma \rightarrow 0$, the coupling vanishes and the minimum of the potential is shifted to $\phi \rightarrow + \infty$, which is consistent with a potential with no minimum value, as was the case for the uncoupled system. This is a reflection of the fact that the scaling solution is allowed to exist only when the effective potential has global minimum values. Also, consistently, the minimum value of the effective potential, presented in Eq.~\eqref{phim}, is valid only when $\lambda>0$ and $\sigma <0$, corresponding to a power-law form for the conformal coupling function [see Eq.~\eqref{cons}]. When $\sigma>0$, the conformal coupling function takes the form of an inverse power law, which has no effective minimum.

\section{Results and Cosmological Analysis} \label{sec:vc}

Finally, we seek viable cosmologies, i.e., trajectories on the phase space capable of reproducing the expansion history of the Universe: start deep in a radiation dominated era, then enter a matter dominated regime, and, finally, evolve towards the scalar field attractor. The stability conditions for k-essence models can be found in Ref.~\cite{amendola}, from which we determine that the tachyonic dark energy model is free from quantum instabilities and superluminal propagation.

The only fixed points depicting radiation and matter dominance in the Universe are fixed points with $y=0$. Naturally, this implies that, in this framework, even though a radiation dominated era is always achieved in the past, the transition to a matter dominated epoch can be successfully achieved only if $y \approx 0$ at early times. Hence, we find that it is possible to recover past radiation and matter dominated eras, leading to the conformally coupled scenario with (D) or (S) as possible future attractors at late times.

As we have seen from the dynamical analysis of the model, the introduction of the coupling is accompanied by the emergence of a scaling solution that, for a specific value of parameters, is able to provide accelerated expansion. Indeed, since in this model the relativistic fluids are not coupled to the scalar field, the deviations of the cosmological evolution in relation to the uncoupled case are evident only after radiation domination. Consequently, the radiation-matter equality epoch does not change significantly in the coupled scenario.

For the uncoupled case ($\sigma = 0$), viable cosmologies correspond to orbits passing close to
\begin{equation*}
(\text{B}_+)\ \text{or}\  (\text{B}_-)\  \text{and}\ (\text{C})\ \longrightarrow   (\text{O})\ \longrightarrow (\text{D}).
\end{equation*}
In this context, the evolution of the cosmological parameters is fully described, according to the information provided in Table~\ref{table:gamma1} and is widely addressed in the literature \cite{Bahamonde:2017ize}. 
The most important trait is definitely the evolution of the energy densities of the fluids. At early times, radiation always dominates, followed by a matter dominated era. The late-time transition from a matter to a dark energy era occurs naturally in the phase space. Even though it is possible to fit the curves in order to recover the present observed value of $\Omega_{\phi} \simeq 0.7$ \cite{planck2}, the Universe will inevitably evolve towards a full dark energy dominated regime.

With the introduction of the coupling ($\sigma \neq 0$), we take the cosmological evolution corresponding to orbits shadowing the transition 
\begin{equation*}
(\text{B}_+)\ \text{or}\ (\text{B}_-)\ \longrightarrow (\text{A}_+)/(\text{A}_-)\ \longrightarrow (\text{D})\ \text{and/or}\  (\text{S}).
\end{equation*}
In this case, there is a new fixed point which is a scaling solution and which can be an attractor for certain regimes. However, for some values of the parameters, this solution presents a spirally attractive nature. This effect is stronger for larger absolute values of $\sigma$ (and constant $\lambda$).

For what concerns the EoS of the tachyon field, as observations indicate that $w_{\phi} \simeq -1$ \cite{planck2}, in this framework, $\dot{\phi}$ must currently be close to zero, requiring either small values of $\lambda$ or large negative values of $\sigma$.
For what concerns the effective EoS parameter, at early times, it must be close to $1/3$ and, as nonrelativistic matter becomes the dominant component in the Universe, it evolves towards $w_{\rm eff}=0$ before approaching the value at the attractor which, accordingly, must resemble $w_{\rm eff} \simeq 0.7$. We will now consider a few examples on how the scalar field dominated and the conformal scaling fixed points can be used to deliver interesting cosmological scenarios. However, in what follows, we cannot neglect the contribution of baryons to the energy density budget, which we ignored as a first approximation. We take this component to be uncoupled to the field, in order to avoid conflicts related to constraints from Solar System bounds.
\begin{figure*}[t]
\subfloat{\includegraphics[width=0.47\linewidth]{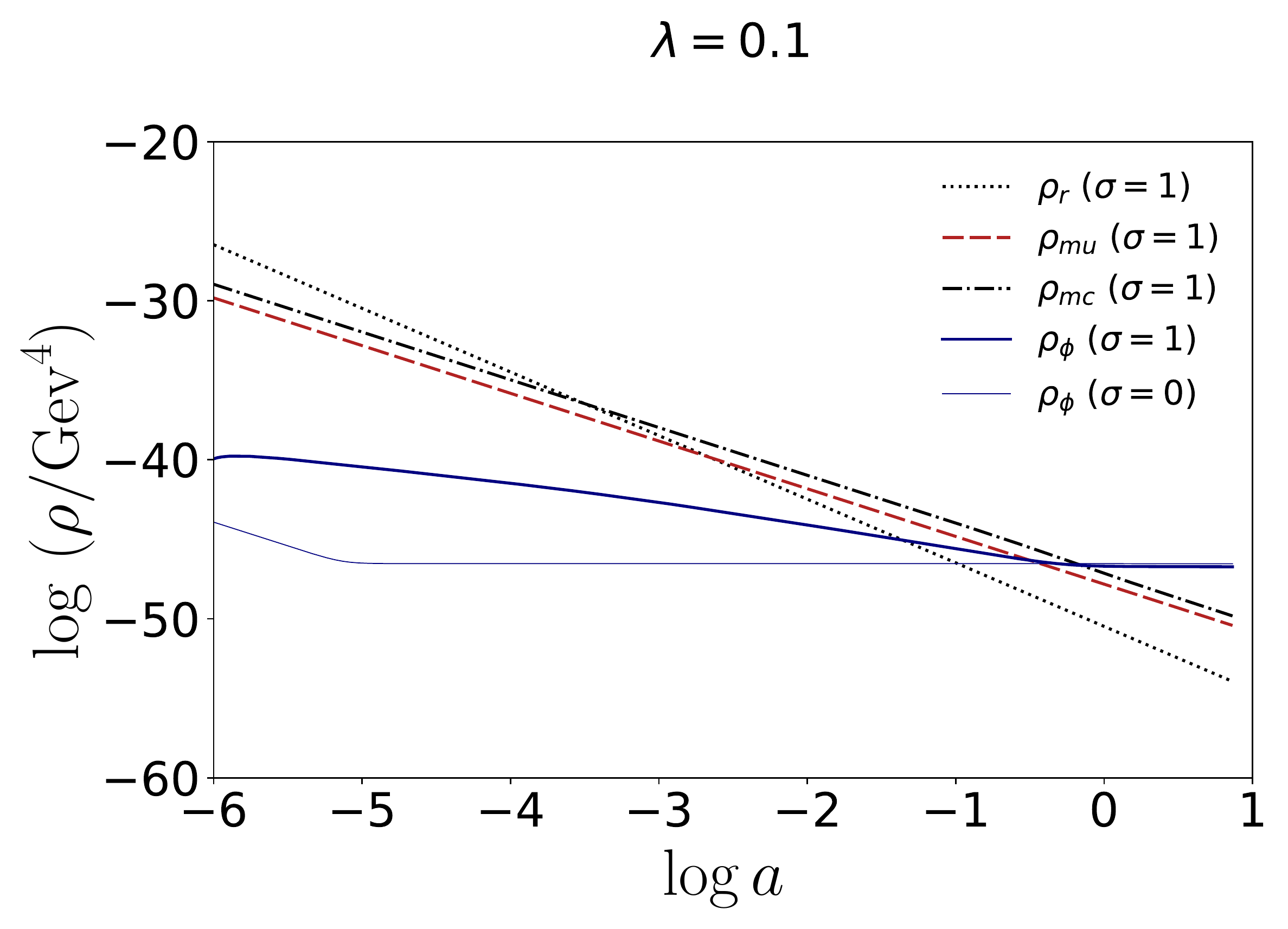}} 
        \hfill
      \subfloat{\includegraphics[width=0.47\linewidth]{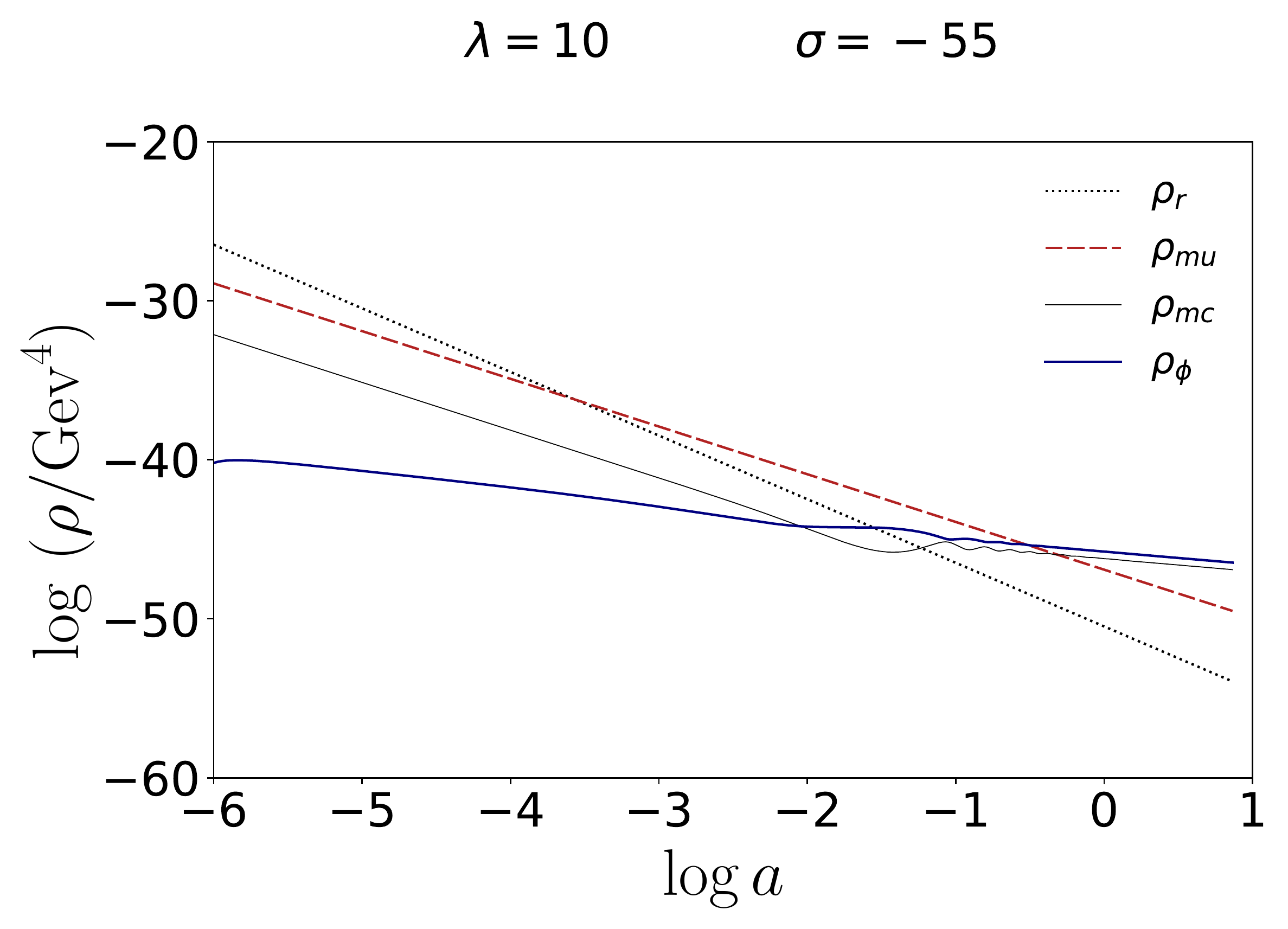}} 
      \hfill
       \subfloat{\includegraphics[width=0.47\linewidth]{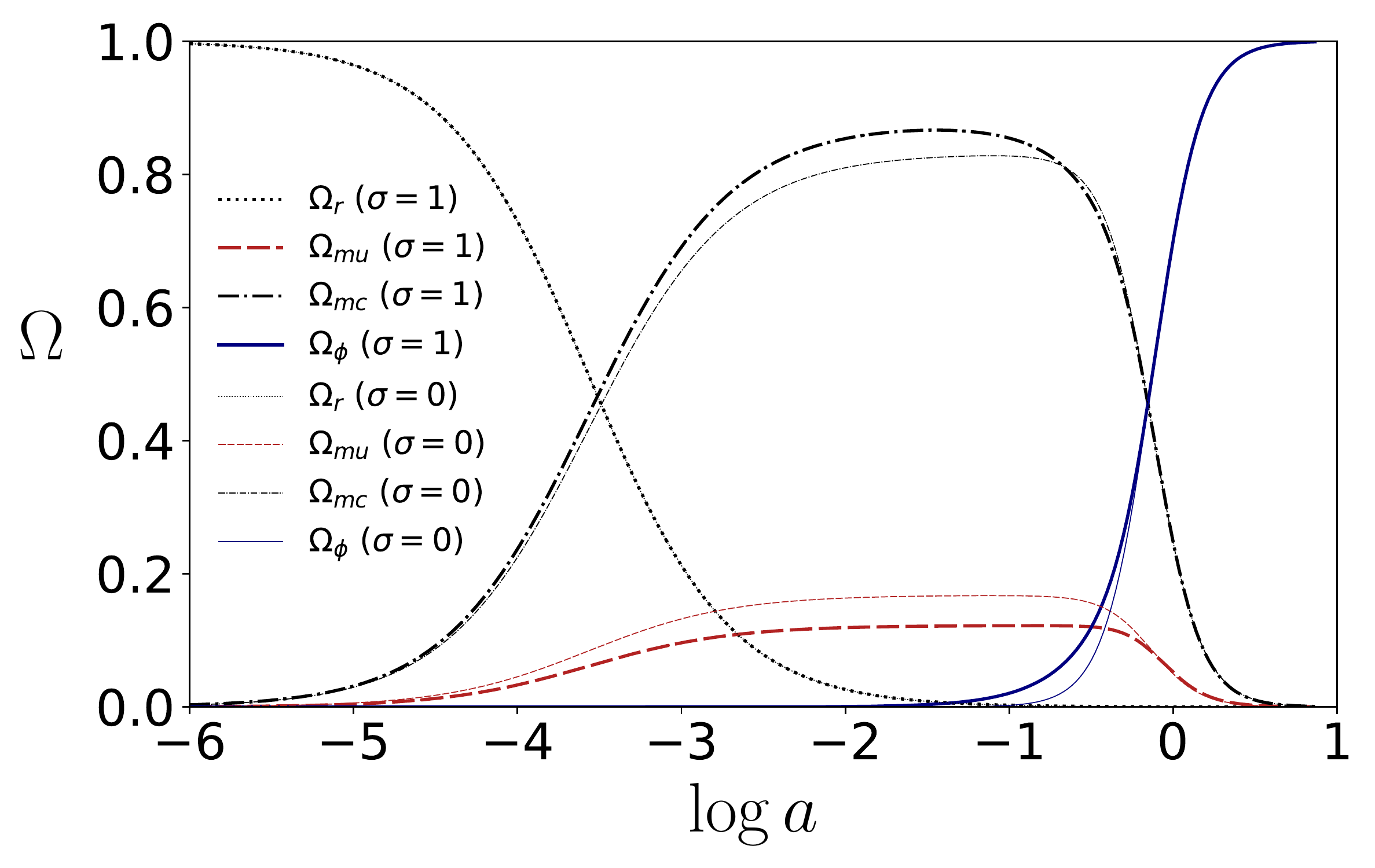}}
   \hfill
       \subfloat{\includegraphics[width=0.47\linewidth]{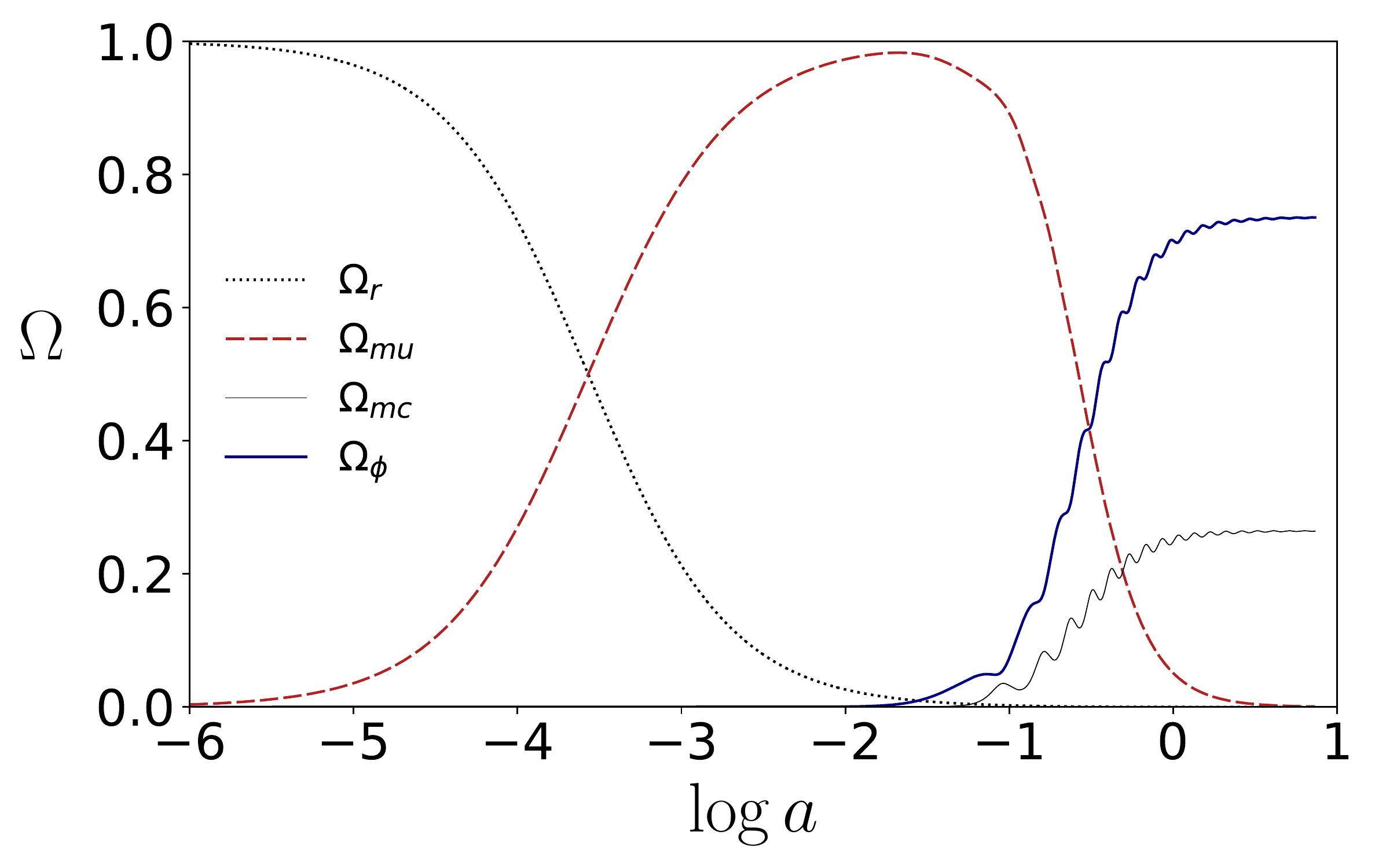}}
       \hfill
    \subfloat{\includegraphics[width=0.47\linewidth]{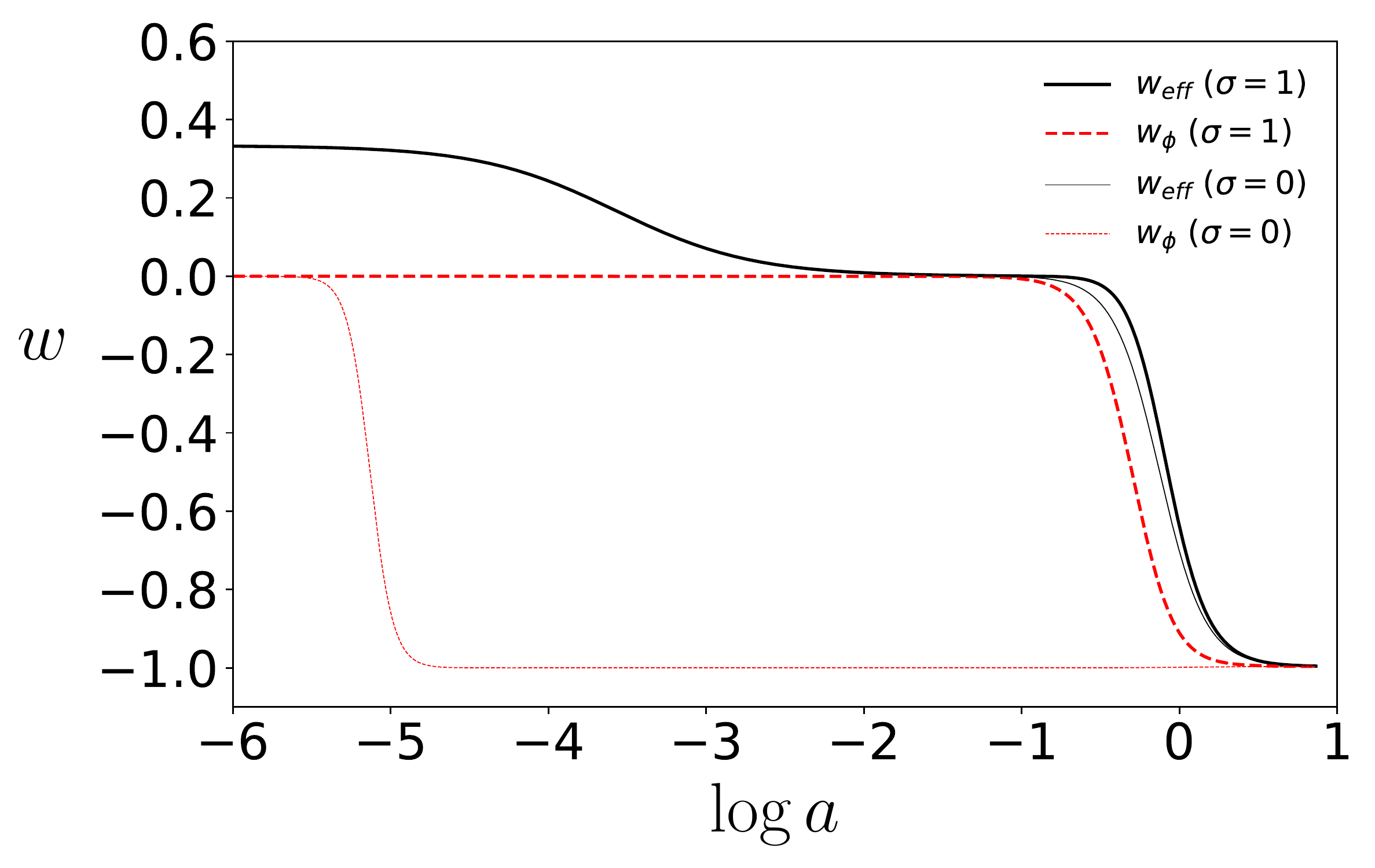}}
    \hfill
    \subfloat{\includegraphics[width=0.47\linewidth]{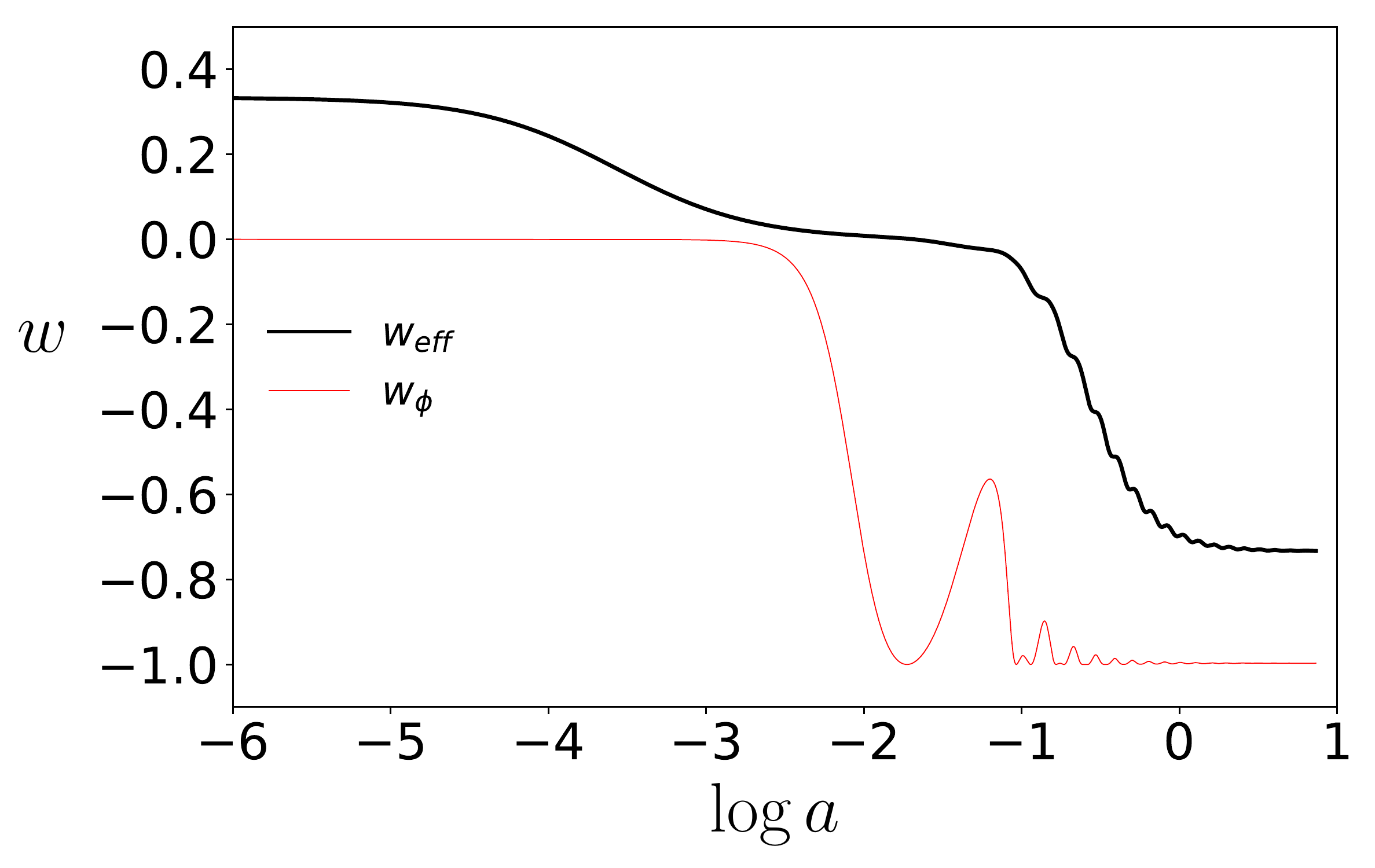}}
  \caption{\label{fig:cosmpar}  Illustration of (top panels) the energy densities, $\rho_{\phi, m,r}$, and (middle panels) density parameters, $\Omega_{\phi,{m},{r}}$, for the field, matter (the subscripts $mu$ and $mc$ distinguish between uncoupled  and coupled matter sources) and radiation, respectively, and of (bottom panels) the EoS parameters, $w_{\rm eff}$ and $w_{\phi}$. (Left panels) Evolution of the cosmological parameters when the attractor is the scalar field dominated solution (D) for $\lambda=0.1$ and $\sigma =1$, in comparison to the $\sigma=0$ case. (Right panels) The attractor is the scaling fixed point (S). At present, we are very close to a Universe evolving under the scaling regime with $\Omega_\phi \approx 0.7$ and $\Omega_mu \approx 0.05$, and the coupling is fixed by $\lambda=10$ and $\sigma =-55$, according to Eq.~\eqref{eqnel}.}
\end{figure*}
\begin{figure*}[t]
\subfloat{\includegraphics[width=0.47\linewidth]{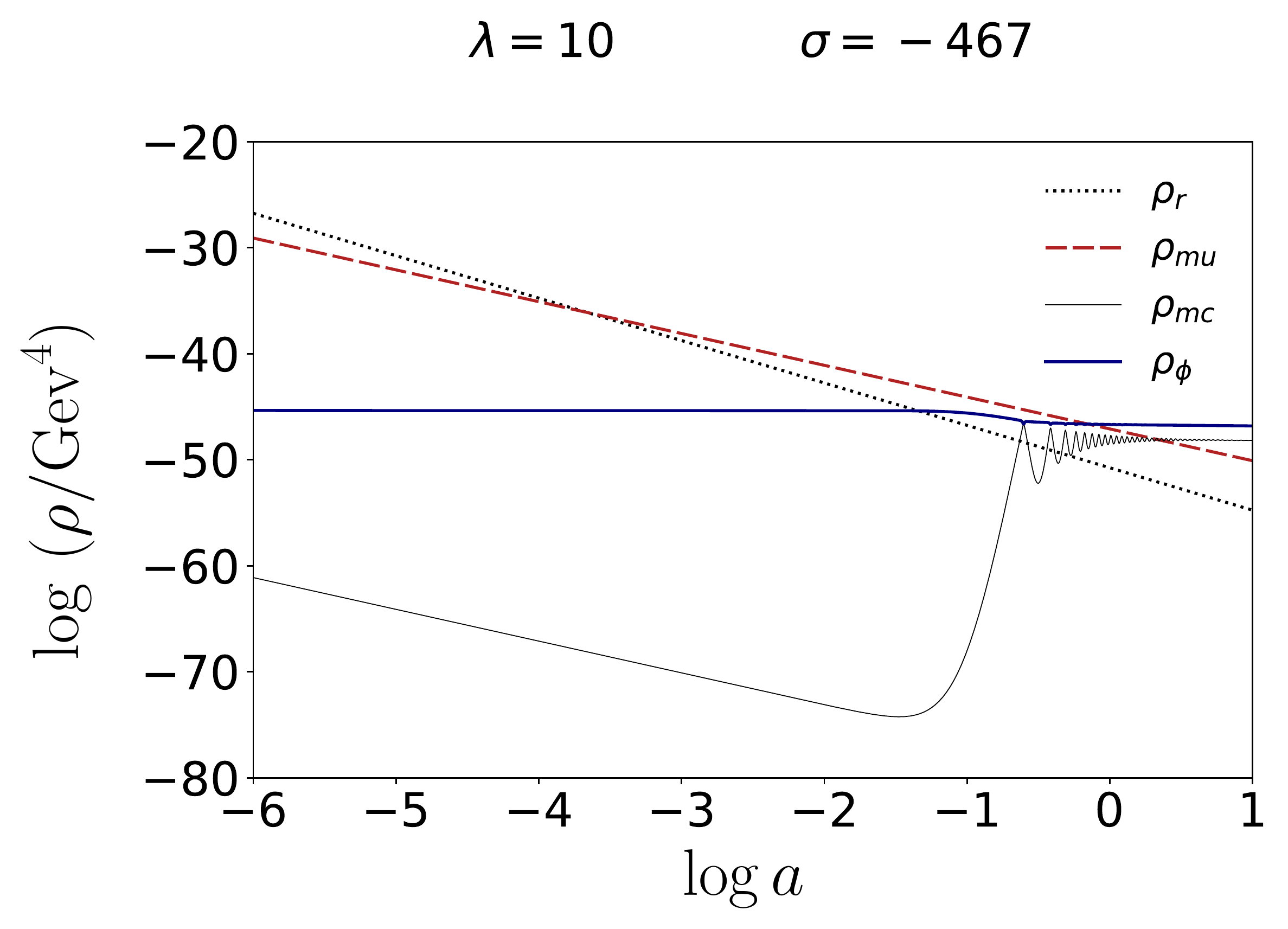}} 
        \hfill
      \subfloat{\includegraphics[width=0.47\linewidth]{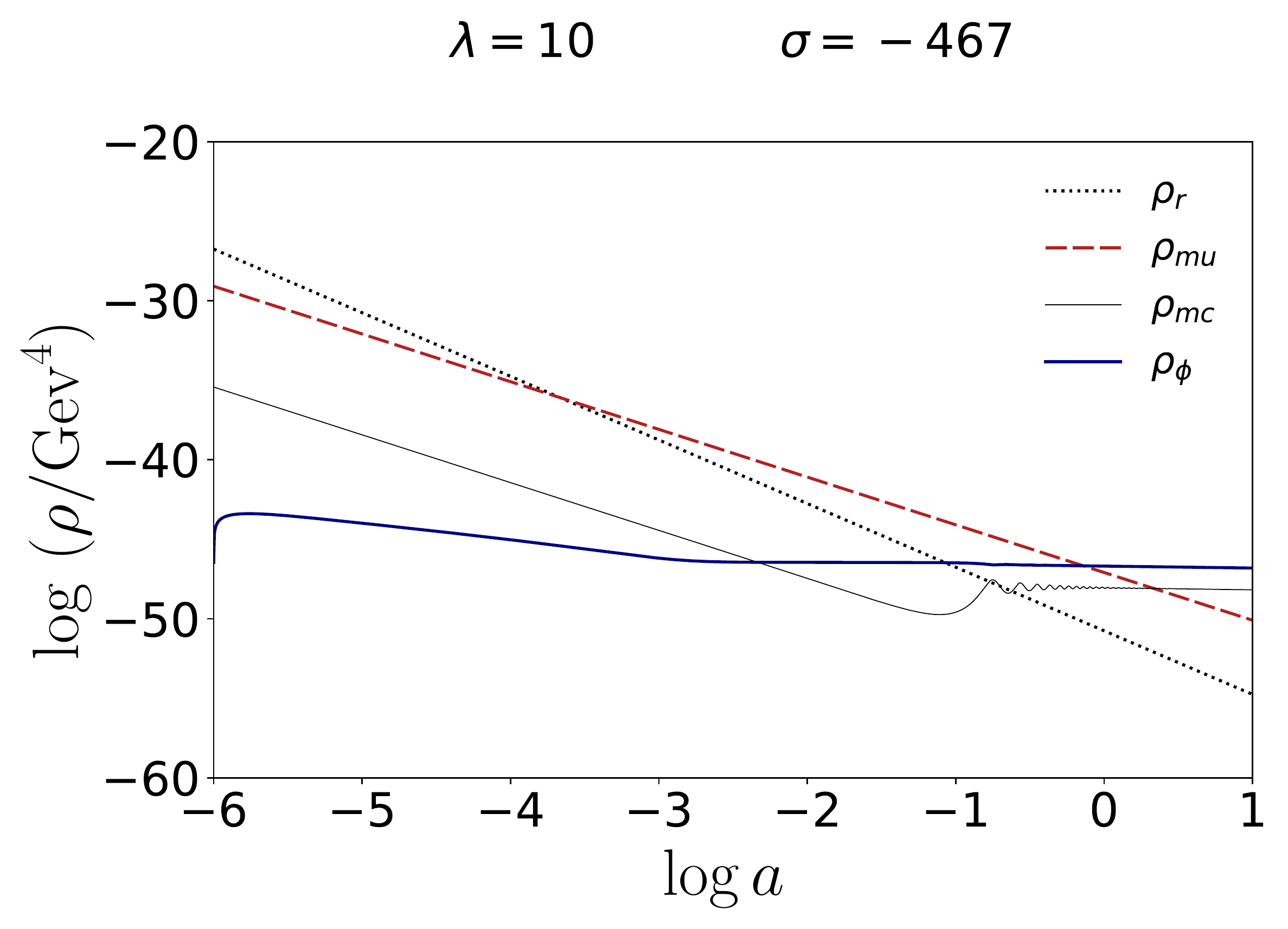}} 
      \hfill
       \subfloat{\includegraphics[width=0.47\linewidth]{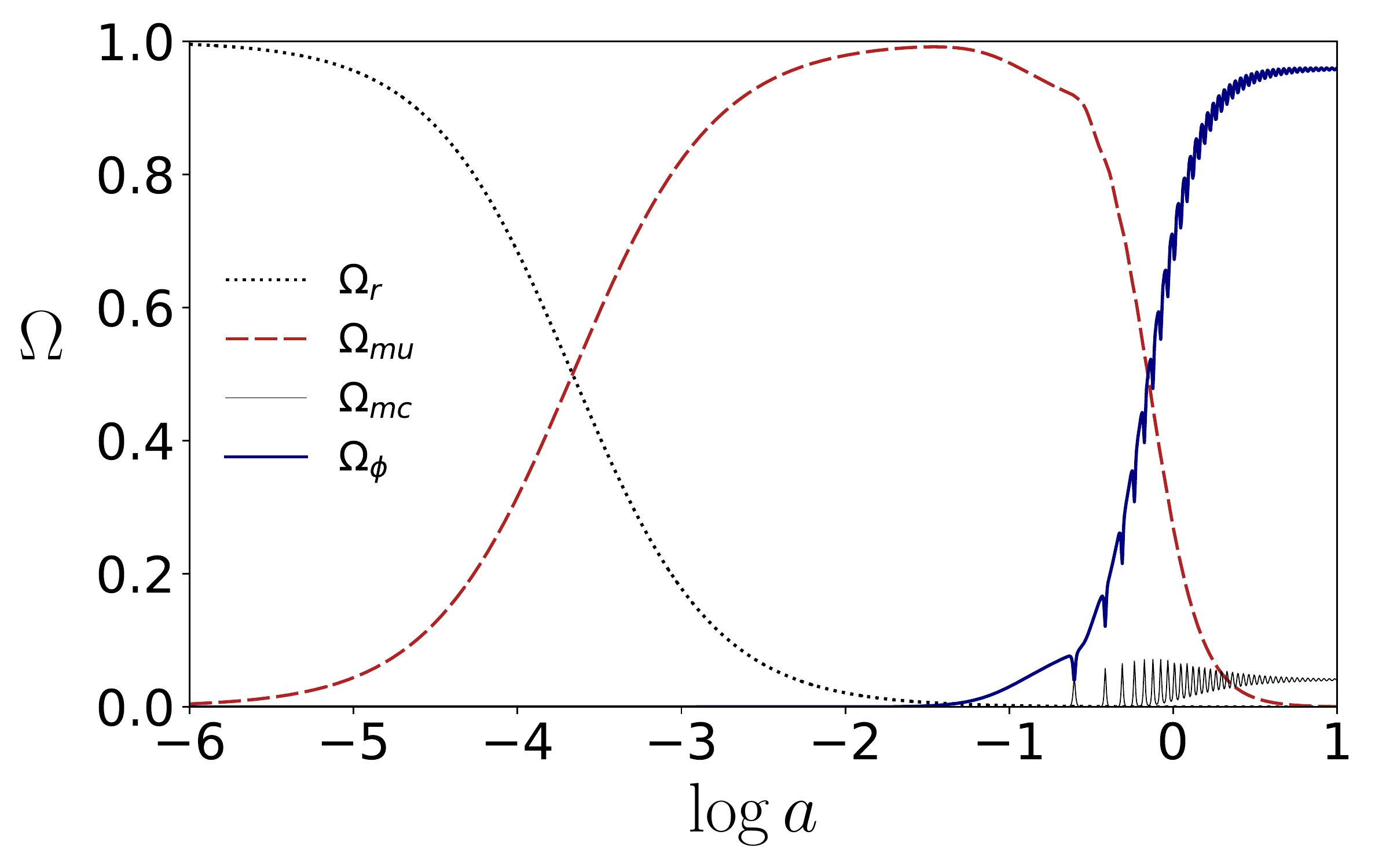}}
   \hfill
       \subfloat{\includegraphics[width=0.47\linewidth]{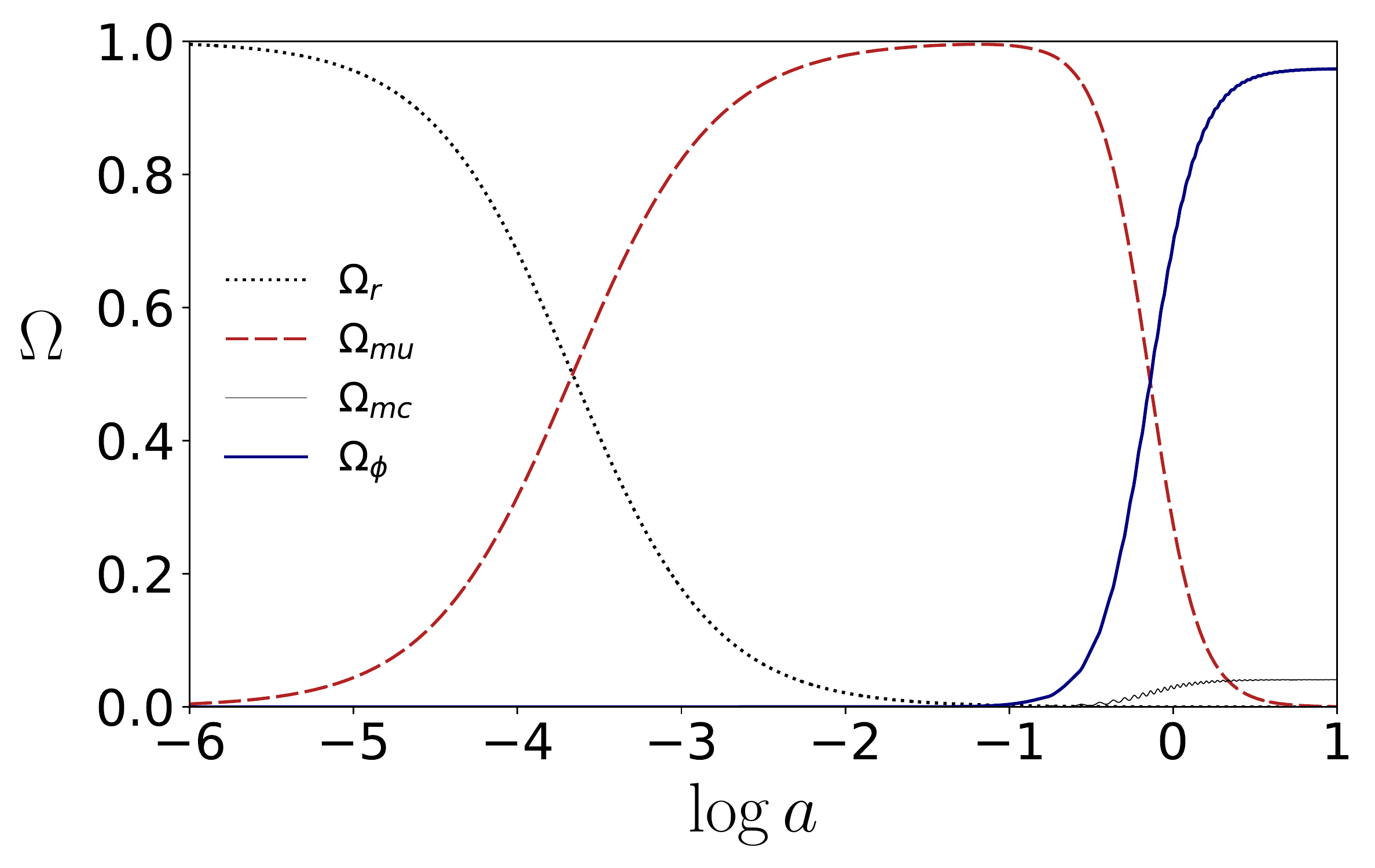}}
       \hfill
    \subfloat{\includegraphics[width=0.47\linewidth]{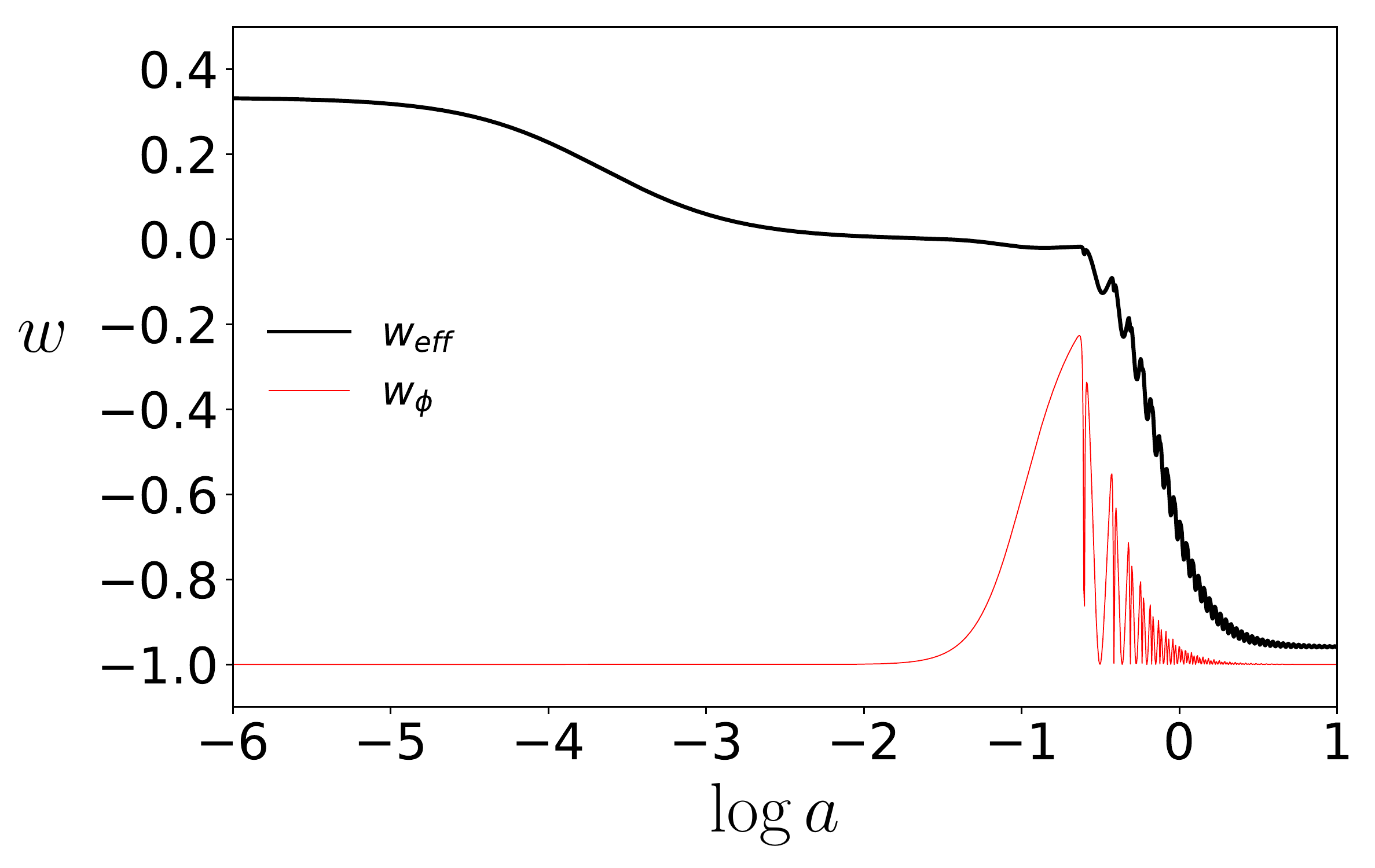}}
    \hfill
    \subfloat{\includegraphics[width=0.47\linewidth]{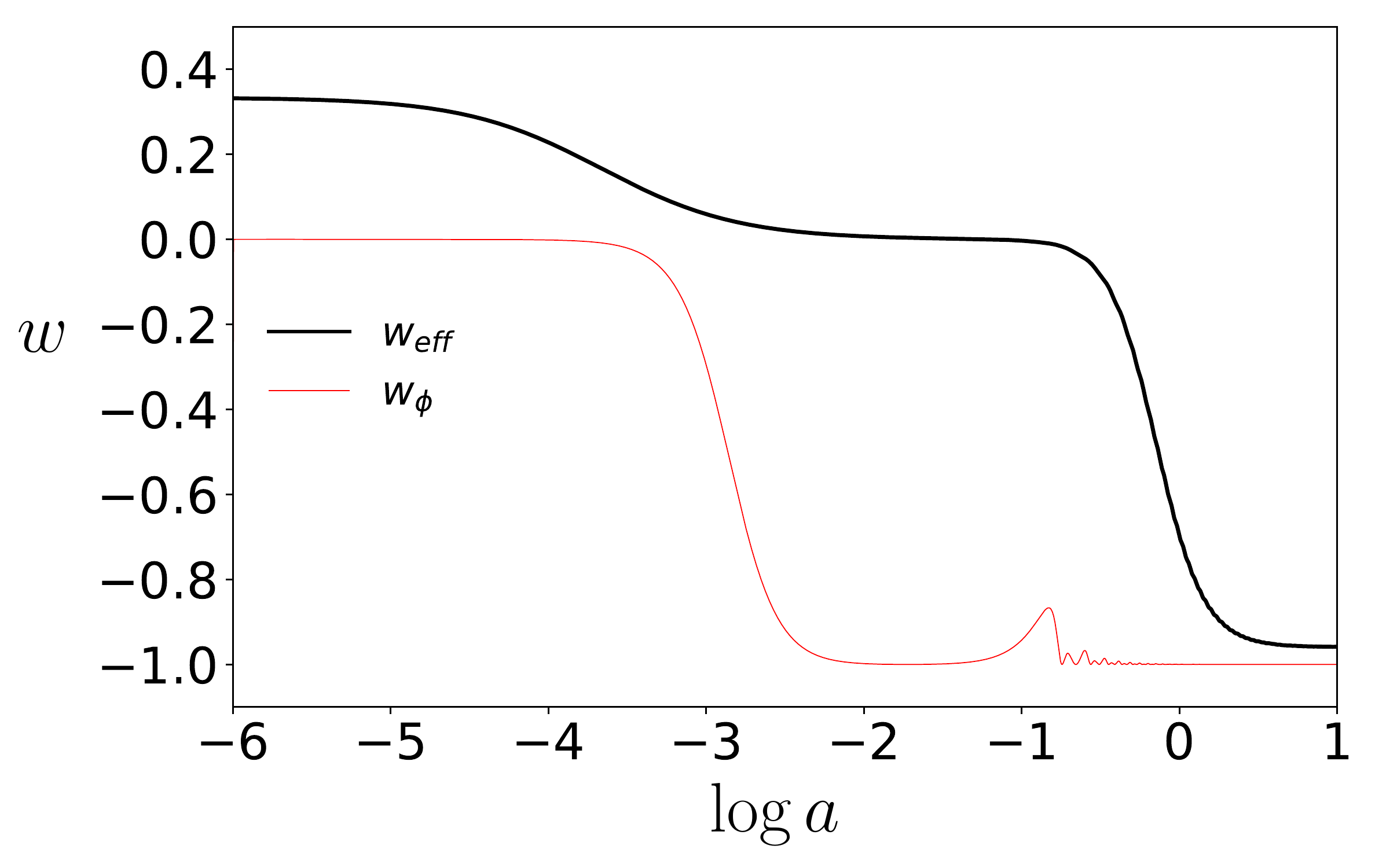}}
  \caption{\label{fig:cosmpar2} Illustration of (top panels, logarithmic scale) the energy densities, $\rho_{\phi, m,r}$, and (middle panels) density parameters, $\Omega_{\phi,{m},{r}}$, for the field, total matter (the subscripts $mu$ and $mc$ stand for uncoupled and coupled matter sources), and radiation, respectively, and (bottom panels) the EoS parameters, $w_{\rm eff}$ and $w_{\phi}$. The attractor of the system is the scaling fixed point (S) and the coupling is fixed through $\lambda=10$ and $\sigma=-467$, according to Eq.~\eqref{eqnel}, with $\Omega_\phi \simeq 0.7$ and $\Omega_{mu} \simeq 0.27$ at present. (Left panels) The initial condition is specified such that the field rolls down the effective potential through the branch where the inverse square potential is dominant, resembling the uncoupled system at early times. (Right panels) The initial condition for $\phi$ is such that the field rolls down the effective potential through the branch where the contribution of the coupling function is dominant.}
\end{figure*}

\subsubsection{The scalar field dominated fixed point}

Without the coupling, the attractor is always the scalar field dominated fixed point which can provide an accelerated expanding scenario for $\lambda^4<12$. When the coupling is present, the fixed point (D) is still a possible attractor of the system (given that $\sigma \geq \widetilde{\lambda}$), which can feature accelerated expanding behaviour for the same conditions.
The differences in relation to the uncoupled case can be seen in the evolution of the EoS parameter of the field. Initially the field behaves as matter [fixed points (B$_\pm$) and (A$_\pm$)], instead of freezing near $w_{\phi} = -1$ [fixed points (C) and (O)], before evolving towards the value at the attractor. 
This trait is also evident in the evolution of the energy density of the field, which resembles a cosmological constantlike evolution for the uncoupled case but depicts a clear dynamical evolution when the coupling is present. 
As for the evolution of the relative energy densities, the presence of the coupling manifests itself by a transfer of energy from the matter (DE) into the DE (matter) fluids for $\sigma >0$ ($\sigma<0$). 

An example of the scalar field dominated regime is illustrated on the left panels of Fig.~\ref{fig:cosmpar} for $\lambda=0.1$ and $\sigma=1$, in comparison with the $\sigma=0$ case. The evolution of the components (except for the energy density of the field) is similar to the standard $\Lambda$CDM model.

\subsubsection{Scaling solution}

On the other hand, if $\sigma \leq \widetilde{\lambda}$, with the emergence of the scaling fixed point (S), it is possible to have an everlasting Universe with $\Omega_{\phi} \simeq 0.7$, therefore alleviating the cosmological coincidence problem. In these conditions, an accelerated expanding scenario can be achieved only for large values of $\sigma$, as we saw in Fig.~\ref{fig:regs}. 
However, if $\sigma$ is large and only a small fraction of the matter sector is in an uncoupled form (baryons), the spiralling behaviour around the attractor leaves a strong oscillating late-time signature on the evolution of the cosmological parameters. We expect that this will inevitably lead to large contributions of the DE field at early times and/or to instabilities in the perturbations. This problem can be eased by considering that most of the matter sector is uncoupled at early times. If we take all of the uncoupled matter to be in the form of baryons, the abundance of dark matter must always be subdominant, which suggests difficulties in the formation of structure in galaxies and clusters of galaxies. A possible way to soothe this apparent problem is to consider that most of the dark matter is also uncoupled and only a small fraction interacts with the scalar field
(see, for example, Refs.~\cite{Copeland:2003cv, Nunes:2011mw, Amendola:2007yx}). 
This scenario is depicted in the right panels of Fig.~\ref{fig:cosmpar}, where the cosmic coincidence problem is alleviated by having $\Omega_{\phi} \simeq 0.7$ at the scaling solution and most of the dark matter component coupled to the field. The scaling solution is almost reached at present, apart from the presence of the baryons in the uncoupled source.

Taking into account the value of $\Omega_\phi$ in Table~\ref{table:gamma1}, an approximate empirical relation for $\sigma$ can be estimated for a specific value of $\lambda$ and given values of $\Omega_\phi$, $\Omega_{mu}$, and $\Omega_{mc}$ at present,
\begin{eqnarray}
\sigma \approx 2\lambda - \frac{\lambda+\sqrt{\lambda^2-12\Omega_{mc}/(\Omega_{mc}+\Omega_\phi)}}{\Omega_{mc}/(\Omega_{mc}+\Omega_\phi)} 
\label{eqnel},
\end{eqnarray}
\noindent valid in the limit where $\lambda$ and $\sigma$ are far larger than unity and where we have written $\Omega_\phi-1$ as $\Omega_{mc}/(\Omega_{mc}+\Omega_\phi)$.

Given the effective scalar field potential considered in Sec. \ref{sec:eff}, it is possible to distinguish between two characteristic cosmic evolutions of the system. The left panel of Fig.~\ref{effpot} shows that the effective potential is not symmetric in $\phi$, and furthermore, it is a clear composition of the effect of the coupling function and the scalar field potential. Therefore, if the initial condition for $\phi$ is given such that the field is rolling down the effective potential through the branch where the scalar field potential is dominant (the left side of the effective potential in the left panel of Fig.~\ref{effpot}), the evolution of the system will resemble the uncoupled regime at early times (characterised by the uncoupled fixed points in Table~\ref{table:gamma1}), before evolving towards the scaling solution, corresponding to the field oscillating in the minimum of the effective potential. 
One example of such an evolution is depicted in the left panels of Fig.~\ref{fig:cosmpar2}, with $\lambda=10$ and the value of $\sigma$ given by Eq.~\eqref{eqnel}. Note how the energy density of the coupled matter species is negligible at early times when compared to the other components.
Analogously, the field could be rolling down the effective potential through the branch where the contribution of the coupling function is dominant (the right side of the effective potential in the left panel of Fig.~\ref{effpot}), leading to a dynamical evolution of the system such as described in the previous sections (characterised by the coupled fixed points in Table~\ref{table:gamma1}). This regime is illustrated in the right panels of Fig.~\ref{fig:cosmpar2} for the same values of $\lambda$ and $\sigma$. In this case, the energy density of the coupled matter species is higher than the energy density of the field at early times, highlighting its important contribution, albeit being subdominant when compared to the uncoupled components.

In Fig.~\ref{fig:cosmpar2}, we take $\Omega_{um} \simeq 0.27$ today. It can be observed that the final configuration of the Universe corresponding to the scaling solution is not yet reached, and, consequently, the observed value of $\Omega_\phi \approx 0.7$ is valid only at the present.

Nonetheless, all of the previous examples concern strong coupling regimes, and typically, in frameworks such as standard coupled quintessence and growing neutrino scenarios, this leads to interactions stronger than gravity itself and consequently to instabilities in the perturbations. This is an issue to be addressed in a future publication.

In Ref.~\cite{Copeland:2006wr}, we find a discussion regarding the range of values of $\lambda$ that allow for viable cosmological solutions given that, for the uncoupled case, the only viable late-time attractor is the scalar field dominated fixed point (D). This solution requires $\lambda^4 <12$ for the accelerated expansion to occur near the fixed point solution, which translates into a constraint on the mass scale of the potential $V_0$: $V_0 \gtrsim 1.1 M_{\rm P}$. The fact that the energy scale of the potential needs to be greater than the Planck mass in order to obtain a significant acceleration can be problematic, as General Relativity is expected to break down at this scale.
This apparent problem can be alleviated with the presence of the scaling solution (S). When that is the case, accelerated expansion can be achieved for $\sigma < - \lambda$, meaning that the value of $\lambda$ is completely free and that there is an imposition only on the value of $\sigma$. Thus, there are no constraints on the mass of the potential, and the effect is shifted to the scale of the coupling function.

\section{Conclusions}
\label{sec:conc}

In this paper, we have analysed a cosmological model, where the role of dark energy is played by a tachyon field $\phi$, coupled to the matter sector by means of a $\phi$-dependent conformal transformation of the metric tensor. We have considered the simplest known case of an inverse square potential associated with the tachyon field, which corresponds to taking $\lambda$, as defined in Eq.~\eqref{variables}, to be a constant. This model has already been extensively studied for the case in which there is no coupling. However, this is not the most general case, as there is no fundamental reason to assume that the species should not interact. 

We have derived the cosmological equations depicting the evolution of the field and the fluids and have performed a detailed dynamical analysis of the system. 
In Eq.~\eqref{variables}, we have introduced a new quantity, $\sigma$, directly proportional to the interaction term, which, for simplicity, we have taken to be a constant. Consistently, when $\sigma=0$, there is no interaction allowed between the fluids. When $\sigma \neq 0$, we gather that there are six fixed points of the system, as presented in Table~\ref{table:gamma1}. Most of them are well-known critical points, including one scalar field dominated fixed point, labelled as (D), which is now a stable attractor of the system for $\sigma \geq \widetilde{\lambda}$, as defined in Eq.~\eqref{theta}. The introduction of the coupling is associated with the emergence of a new fixed point, labelled as (S), a potentially cosmological viable scaling solution which exists and is a stable attractor for $\sigma \leq \widetilde{\lambda}$ and, moreover, can provide accelerated expansion for 
$\sigma<-\lambda$. It is clear that the fixed points (D) and (S) are related through a transcritical bifurcation in the parameter space. 
We conclude that there are two potential attractors of the system, characterised by a totally disjoint parameter region, meaning that there is always one and only one late-time attractor solution.

Additionally, we have performed a detailed analysis of the cosmological evolution of the tachyonic coupled system in two regimes. First, when the attractor solution is the fixed point (D), a direct comparison with the uncoupled case can be made. The introduction of the coupling manifests itself through a transfer of energy from the matter (DE) into the DE (matter) fluids for $\sigma >0$ ($\sigma<0$), and viable cosmologies correspond to small values of $\lambda$, as $\lambda^4<12$ is needed for accelerated configurations. Second, the scenarios where the late-time attractor is (S) is of great physical interest, as the scaling solution can be used to alleviate the cosmic coincidence problem, by consideration of an accelerated expanding scenario with $\Omega_{\phi}\simeq 0.7$ and $w_{\rm eff} \simeq -0.7$ today. From the dynamical analysis performed, we gather that, for certain values of the parameters, the spirally attractive nature of the scaling solution leads to oscillations in the energy densities, in which case the model can present problems at the level of the background and, possibly, at the level of the cosmological perturbations. This effect can be soothed with the introduction of an uncoupled matter fluid, dominant at early times, composed by baryons and uncoupled dark matter. We have seen how the different evolutions of the cosmological parameters in this framework can be interpreted by means of an effective potential associated with the scalar field, comprising the effect of the scalar field potential and the coupling function.

The fact that the EoS parameter for the tachyon field always varies between $0$ and $-1$ guarantees that phantom solutions are never present. We have found that the cosmological history of the Universe can only be reproduced if $y\approx 0$ at early times, since past radiation and matter domination eras are achieved by means of fixed points with $y=0$. 

In order to further address the question of viable cosmologies, we need to dive into perturbation theory. This is work to be carried out in the future in order to restrict this model according to the observational constraints.
Additionally, a more general analysis could be performed for the case where $\lambda$ and $\sigma$ are no longer constant, i.e., the potential \cite{cop} and the coupling function are left as free functions of the model. 

Finally, we conclude that the conformally coupled tachyonic system is of great cosmological interest in the sense that it provides a wide variety of physically distinct cosmological evolutions and scenarios. Most importantly, we have found a way of alleviating the cosmic coincidence problem while keeping the most attractive features associated with the tachyon scalar field.

\acknowledgments 
The authors thank Bruno Barros for the assistance with the figures. N. J. N. and E. M. T. acknowledge the financial support by Funda\c{c}\~ao para a Ci\^encia e a Tecnologia (FCT) through the Research Grant No. UID/FIS/04434/2013, by Projects No. PTDC/FIS- OUT/29048/2017, No. COMPETE2020: POCI-01-0145-FEDER-028987, No. FCT: PTDC/FIS-AST/28987/2017, and No. IF/ 00852/2015 of FCT. E. M. T. was supported by the Grant No. BI-MESTRE-IF/00852/2015 from FCT.

\begin{appendices}

\section{Linear Stability Matrix}\label{app:mat}

The stability analysis presented in Sec. \ref{sec:fp}, for the fixed points listed in Table~\ref{table:gamma1}, is made through consideration of the eigenvalues of the $3 \times 3$ linear stability matrix $\mathcal{M}$, evaluated at each fixed point (the eigenvalues can be found in Appendix~\ref{app:eig}). For the case of a pressureless fluid, with $x'$, $y'$, and $r'$ as defined in Eqs.~\eqref{xl})--\eqref{rl}, the matrix elements $\mathcal{M}_{ij}$ are\\

$\mathcal{M}_{11}=\frac{\partial x'}{\partial x} = \frac{3 \left(r^2-1\right) \sigma  \sqrt{3-3 x^2} x}{2 y}+9 x^2+\sqrt{3} x y (\sigma -2 \lambda )-3 $,

$\mathcal{M}_{12}=\frac{\partial x'}{\partial y} = -\frac{\sqrt{3} \left(x^2-1\right) \left(\left(r^2-1\right) \sigma  \sqrt{1-x^2}+y^2 (2 \lambda -\sigma )\right)}{2 y^2}$,

$\mathcal{M}_{13}=\frac{\partial x'}{\partial r} = \frac{r \sigma  \sqrt{3-3 x^2} \left(x^2-1\right)}{y}$,

$\mathcal{M}_{21}=\frac{\partial y'}{\partial x} = -\frac{1}{2} y^2 \left(\sqrt{3} \lambda -\frac{3 x y}{\sqrt{1-x^2}}\right)$,

$\mathcal{M}_{22}=\frac{\partial y'}{\partial y} = \frac{1}{2} \left(r^2-9 \sqrt{1-x^2} y^2-2 \sqrt{3} \lambda  x y+3\right)$,

$\mathcal{M}_{23}=\frac{\partial y'}{\partial r} = r y$,

$\mathcal{M}_{31}=\frac{\partial r'}{\partial x} = \frac{3 r x y^2}{2 \sqrt{1-x^2}} $,

$\mathcal{M}_{32}=\frac{\partial r'}{\partial y} = -3 r \sqrt{1-x^2} y$,

$\mathcal{M}_{33}=\frac{\partial r'}{\partial r} = \frac{1}{2} \left(3 r^2-3 \sqrt{1-x^2} y^2-1\right)$. \\

\section{Eigenvalues of the Stability Matrix}
\label{app:eig}

Here we present the eigenvalues of the matrix $\mathcal{M}$, as defined in Appendix~\ref{app:mat}, evaluated at each fixed point, as presented in Table~\ref{table:gamma1}. The eigenvalues are used to infer the stability character of each fixed point. The eigenvalues for the fixed point (S) are not presented due to their extensive expression.

\begin{itemize}

\item Fixed point (O) ($\sigma \equiv 0$):

$e_1=-3$, $e_2=\frac{3}{2}$, $e_3=-\frac{1}{2}$.

\item Fixed points (A$_\pm$) ($\sigma \equiv 0$):

$e_1=6$, $e_2=\frac{3}{2}$, $e_3=-\frac{1}{2}$.

\item Fixed points (B$_\pm$):

$e_1=6$, $e_2=2$, $e_3=1$.

\item Fixed point (C) ($\sigma \equiv 0$):

$e_1=-3$, $e_2=2$, $e_3=1$.

\item Fixed point (D):

$e_1=\frac{1}{12} \left(-\lambda ^4+\sqrt{\lambda ^4+36} \lambda ^2-36\right)$,

$e_2=\frac{1}{12} \left(-\lambda ^4+\sqrt{\lambda ^4+36} \lambda ^2-24\right)$,

$e_3=\frac{1}{12} \lambda  \left(\sqrt{\lambda ^4+36}-\lambda ^2\right) (2 \lambda -\sigma )-3$.\\

\end{itemize}

\section{Bifurcation} \label{sec:mybif}

As addressed in Sec. \ref{sec:fp}, the conformally coupled system features one bifurcation involving the fixed points (D) and (S). 
Through inspection of the eigenvalues of the stability matrix, evaluated at the fixed point (D) (found in Appendix~\ref{app:eig}), we infer that its stability character depends only on the value of $e_3$:
\begin{equation}
e_3=\frac{1}{12} \lambda  \left(\sqrt{\lambda ^4+36}-\lambda ^2\right) (2 \lambda -\sigma )-3.
\end{equation} 
\noindent The remaining eigenvalues are independent of $\sigma$ and are negative $\forall\  \lambda \in \mathds{R}$. Hence, if $e_3<0$, the fixed point (D) is an attractor, whereas if $e_3>0$, it is a saddle. Accordingly, its stability character changes when
\begin{equation}
e_3=0 \Longrightarrow \sigma= \lambda -\sqrt{\frac{\lambda ^4+36}{\lambda ^2}},
\label{sigb}
\end{equation}
\noindent which, according to the local analysis performed in Sec \ref{sec:fp} for each fixed point, coincides with the value of $\sigma$ for which the fixed point (S) enters the phase space, as defined in Eq.~\eqref{l1}. This, together with a proper analysis of the numerical behaviour of the system, allows us to recognise that the fixed points (D) and (S) undergo a transcritical bifurcation when $\sigma$ takes the value in Eq.~\eqref{sigb} (for each specific value of $\lambda$), which corresponds to the bifurcation point and was verified numerically. 

\end{appendices}
%\bibliographystyle{unsrt}

% External bibliography database file in the BibTeX format
\bibliography{mybib} 

\end{document}